# C to Checked C by 3C*

March 25, 2022

ARAVIND MACHIRY, Purdue University, USA
JOHN KASTNER, Amazon, USA
MATT MCCUTCHEN, Amazon, USA
AARON ELINE, Amazon, USA
KYLE HEADLEY, Amazon, USA
MICHAEL HICKS, Amazon, USA

Owing to the continued use of C (and C++), spatial safety violations (e.g., buffer overflows) still constitute one of today's most dangerous and prevalent security vulnerabilities. To combat these violations, *Checked C* extends C with bounds-enforced *checked pointer* types. Checked C is essentially a *gradually typed* spatially safe C—checked pointers are backwards-binary compatible with legacy pointers, and the language allows them to be added piecemeal, rather than necessarily all at once, so that safety retrofitting can be incremental.

This paper presents a semi-automated process for porting a legacy C program to Checked C. The process centers on **3C**, a static analysis-based annotation tool. **3C** employs two novel static analysis algorithms—**typ3c** and **boun3c**—to annotate legacy pointers as checked pointers, and to infer array bounds annotations for pointers that need them. **3C** performs a *root cause analysis* to direct a human developer to code that should be refactored; once done, **3C** can be re-run to infer further annotations (and updated root causes). Experiments on 11 programs totaling 319KLoC show **3C** to be effective at inferring checked pointer types, and experience with previously and newly ported code finds **3C** works well when combined with human-driven refactoring.

## 1 INTRODUCTION

Vulnerabilities due to memory corruption are still a major issue for C programs [Trends 2021] despite a large body of work that tries to prevent them [Song et al. 2019]. Microsoft's 2019 Blue Hat analysis [BlueHat 2019] found that spatial memory safety issues (*invalid memory access*es such as buffer overflows) were the most common vulnerability category; MITRE's CWE top-25 list for 2021 [MITRE 2021] ranks out-of-bounds reads/writes as two of its top three.

Prior tools [Szekeres et al. 2013], including CCured [Necula et al. 2005], Softbound [Nagarakatte et al. 2009], Low Fat pointers [Duck and Yap 2016], and Address Sanitizer (ASAN) [Serebryany et al. 2012] aim to enforce spatial safety automatically, by analyzing a C program and compiling it to include run-time safety checks. Unfortunately, the resulting run-time overhead is too high for deployment (between 60%-200%), and gaps in what programming idioms a tool can handle either cause some programs to be rejected or leave portions of them unprotected. Rather than compile in safety checks directly, a tool could convert C to a memory-safe language that has them, e.g., Rust, a promising memory-safe systems language [Mozilla 2021]. Unfortunately Rust is too different from C to constitute a practical target [Zeng and Crichton 2019], at least for now: the (best of breed) `c2rust` tool [c2r 2021; Larson 2018] transliterates C to *unsafe, non-idiomatic* Rust, and recent work [Emre et al. 2021] is able to take only small steps to close the safety gap.

Recently, Tarditi et al. [2018] proposed *Checked C* as a practical target to which to convert legacy/active C code. Checked C extends C with *checked pointer types* which are restricted by the

*This work was carried out prior to the authors joining Amazon.

Authors' addresses: Aravind Machiry, Purdue University, USA, amachiry@purdue.edu; John Kastner, Amazon, USA, jkastner@amazon.com; Matt McCutchen, Amazon, USA, matmccut@amazon.com; Aaron Eline, Amazon, USA, aeline@amazon.com; Kyle Headley, Amazon, USA, khead@amazon.com; Michael Hicks, Amazon, USA, mwhicks@amazon.com.

compiler to spatially safe uses.[1] Such pointers have one of three possible types, `ptr`<*T*>, `array_ptr`<*T*>, or `nt_array_ptr`<*T*> (*ptr*, *arr*, and *ntarr* for short), representing a pointer to a single element, array of elements, or null-terminated array of elements of type *T*, respectively. The latter two have an associated *bounds annotation*; e.g., a declaration `array_ptr<int> p : count(n)` says that `p` is a pointer to an `int` array whose size is `n`. Checked C's Clang/LLVM-based compiler represents checked pointers as system-level memory words, i.e., without "fattening" metadata, ensuring backward compatibility. The compiler uses these bounds annotations to add dynamic checks prior to checked pointer accesses, to prevent spatial safety violations. These run-time checks can often be proved redundant and removed by LLVM, yielding good performance. Tarditi et al. reported average run-time overheads of 8.6% on a small benchmark suite, and Duan et al. [2020] found essentially no overhead when running Checked C-converted portions of the FreeBSD kernel.

Could we conceivably carry out an automatic port from C to Checked C that achieves full spatial safety? Probably not: The problems prior tools experienced would remanifest in the emitted Checked C code. However, Checked C's design aims to support *incremental* conversion from legacy C, in the style of *gradual* (aka *migratory*) typing [Greenman and Felleisen 2018; Siek and Taha 2007; Tobin-Hochstadt et al. 2017]. In particular, Checked C permits annotating *some* of a function's parameters or variables as checked pointers, which then benefit from safety checks, while leaving other pointers unannotated. Checked C also supports annotating whole regions of legacy code (e.g., standard libraries) with *interop types* which leave the code alone but provide it with a checked-type interface. Designated *checked regions* of code that use only checked pointers enjoy spatial safety—any run-time spatial safety violation *cannot be blamed* on code in the region [Li et al. 2022; Ruef et al. 2019]. With these mechanisms, an automated conversion tool need not be perfect: It can output partially annotated code, and the developer can take care of parts the tool cannot handle.

This work considers the problem of semi-automatically porting an existing C program to Checked C. We observe that any porting process will essentially involve both *refactoring* code to use idioms that Checked C accepts, and *annotating* that code with Checked C pointer types. We have developed a tool called **3C** (*Checked C Converter*) which automates the annotation part, and guides the developer to the parts of her code that should be refactored. We have developed a two-phase workflow that intersperses runs of **3C** with manual refactoring (Section 3).

**3C** uses two novel static analysis algorithms, which we call **typ3c** (pronounced "Types") and **boun3c** ("Bounce"). **typ3c** runs first to determine which legacy pointers can be made into checked pointers, and then **boun3c** infers the bounds of the *arr* and *ntarr* varieties.

**typ3c** is a whole-program, constraint-based static analysis that works in two parts (Section 4). The first part determines which pointers can be checked (e.g., those not involved in unsafe casts or complex preprocessor uses), and the second determines each checked pointer's type; e.g., pointers `a` that are indexed or incremented (`a[i]` or `a++`) will be *arr* or *ntarr*. **typ3c** is inspired by *type qualifier inference* [Foster et al. 2006], but employs novel components that aim to convert as many pointers as possible to their final, correct checked type. It does this by *localizing wildness* (Section 4.1): When a function `foo` locally uses a pointer parameter unsafely, **typ3c** gives it a Checked C interop type to prevent callers or callees that are otherwise perfectly safe from being polluted by `foo`'s locally unsafe pointer use. This means that when the programmer annotates/refactors `foo` to render it safe, she does not have to also manually update all of its callers/callees (and theirs). **typ3c** helps organize the porting process by identifying the set of *root causes* of non-checked-ness and listing them according to downstream influence: refactoring or annotating the code at the top of the list and then rerunning **3C** will result in converting additional downstream pointers (Section 4.2). **typ3c** employs a novel, multi-step constraint solving algorithm that achieves more general, maintainable results, especially for libraries (Section 4.3).

[1] Checked C does not yet ensure *temporal memory safety*, which means that it does not prevent use-after-free errors. While spatial safety is still quite useful on its own, an extension to Checked C to ensure temporal safety, similar in spirit to CETS [Nagarakatte et al. 2010], is underway [Zhou 2021]. Temporal safety can also be ensured by linking a conservative garbage collector [Boehm and Weiser 1988].

**boun3c** next infers bounds annotations for *arr* and *ntarr* pointers (Section 5); as far as we are aware, **boun3c** is the first analysis that can infer pointer bounds in terms of in-scope variables and constants. **boun3c** takes inspiration from static analyses for race detection, which *correlate* pointers with protective mutex variables [Pratikakis et al. 2011]. **boun3c** instead correlates array pointers with potential bounds employed consistently at pointers' allocation and usage sites. To start, **boun3c** *seeds* bounds at allocation sites; e.g., `int x[10]` seeds `x`'s bound as `count(10)` and `y = malloc(sizeof(int)*n)` seeds `y`'s bound as `count(n)`. These bounds are then propagated consistently via dataflow across scopes; e.g., a call `f(y,n)` propagates `y`'s correlation with `n` to a correlation between `f`'s two parameters. Bounds propagation is treated context-sensitively both for function calls and for `struct`s. When seed bounds are unavailable, **boun3c** tries various heuristics.

**typ3c** aims to be *sound* in the sense that it outputs code that the Checked C compiler will accept so long as **boun3c**'s inferred bounds are correct (modulo bugs in **typ3c** or the Checked C compiler). **boun3c** aims to be *partially sound* in the sense that its (non heuristically) determined bounds are correct, but some may be missing (and again: bugs). We find this balance minimizes porting effort.

**3C** is implemented as a `clang` tool. We evaluated its effectiveness on a benchmark of 11 programs, many of them large, totaling about 371K LoC (Section 6). Running times were fast enough for interactive use—typically less than one second, and at most 22s. Our experiments show that **3C** is effective at inferring Checked C types along with bounds annotations. In particular, we find that **typ3c** automatically converted 67.9% of pointers in our benchmark programs to checked types, which improves on the 48.4% inferred by unification-style algorithms used in prior work [Necula et al. 2005]. **boun3c** was able to infer bounds for 77.3% of pointers that required them. Running **3C** on programs previously ported to Checked C, but with annotations removed, often restored most of the removed annotations, and many times *restored them all*. We have also used **3C** within our two-phase workflow to iteratively port three server programs—`vsftpd`, `thttpd`, and `icecast`—and a bignum library to Checked C, for a total of about 42 KLoC. This process balanced manual and automated work, and using it revealed a (known) CVE in `thttpd` and two new spatial safety violations in the bignum library.

In addition to being different from prior work already mentioned which aims to retrofit C code to be safe, our work represents a novel take on the *automated type migration* problem for gradual typing, which also seeks to automatically infer safety-enhancing static annotations [Phipps-Costin et al. 2021]. Our work differs from all of the above by focusing not on a single automated step, but rather on the iterative process of conversion which leverages automation (Section 7). This usage mode affects the design of the automation: in our case, it affects **3C**'s determination of which pointers are checked, and how it infers array bounds.

**3C** and the programs we have ported with it are freely available as part of the Secure Software Development Project (SSDP)'s fork of Checked C, at https://github.com/secure-sw-dev.

## 2 BACKGROUND: CHECKED C

This section presents some background on Checked C, the target language for **3C**. Checked C [Specification 2016; Tarditi et al. 2018] extends C with support for *checked pointers*. Development of Checked C was initiated by Microsoft Research in 2015 but starting in late 2021 was forked and is now actively managed by the Secure Software Development Project, https://github.com/secure-sw-dev/.

```
1   void baz(int *q,
2            int *c, int len) {
3     for (int i = 0; i<len; i++) {
4       q[i] += *c;
5     }
6   }
7
8   extern void recordptr(void *x);
9
10  static int *g = 0;
11
12  void foo(int *p, int n) {
13
14
15    int m = 0;
16    recordptr(p);
17    g = p;
18    baz(p,&m,n);
19
20  }
21
22  void bar(int z) {
23    int *r =
24      malloc(sizeof(int)*z);
25    foo(r,z);
26    baz(r,g,z);
27  }
```

(a) Original C code

```
void baz(array_ptr<int> q: count(len),
         ptr<int> c, int len) checked {
  for (int i = 0; i<len; i++) {
    q[i] += *c;
  }
}

⚠ extern void recordptr(void *x);

static int *g = 0;

void foo(int *p: itype(array_ptr<int>)
                 count(n),
         int n) {
  int m = 0;
  recordptr(p);
  g = p;
  baz(assume_bounds_cast<array_ptr<int>>
      (p, count(n)),&m,n);
}

void bar(int z) {
  array_ptr<int> r: count(z) =
    malloc<int>(sizeof(int)*z);
  foo(r,z);
  baz(r,assume_bounds_cast<ptr<int>>(g),z);
}
```

(b) After initial conversion with marked root causes (⚠)

```
void baz(array_ptr<int> q: count(len),
         ptr<int> c, int len) checked {
  for (int i = 0; i<len; i++) {
    q[i] += *c;
  }
}

itype_for_any(T) extern void
  recordptr(void *x : itype(array_ptr<T>));
static ptr<int> g = 0;

void foo(array_ptr<int> p: count(n),
         int n) checked {

  int m = 0;
  recordptr<int>(p);
  g = p;
  baz(p,&m,n);

}

void bar(int z) checked {
  array_ptr<int> r: count(z) =
    malloc<int>(sizeof(int)*z);
  foo(r,z);
  baz(r,g,z);
}
```

(c) Complete conversion by **3C** after manually fixing root causes.

Listing 1. (Contrived) Example demonstrating various phases of **3C**.

## 2.1 Checked Pointer Types

Checked pointer types include `ptr<`$T$`>` (*ptr*), `array_ptr<`$T$`>` (*arr*), and `nt_array_ptr<`$T$`>` (*ntarr*), which describe pointers to a single element, an array of elements, and a null-terminated array of elements of type $T$, respectively. Both *arr* and *ntarr* pointers have an associated *bounds* which defines the range of memory referenced by the pointer. Here are the three different ways to specify the bounds for a pointer `p`; the corresponding memory region is at the right.

| | |
|---|---|
| `array_ptr<`$T$`> p: count(`$n$`)` | $[\mathtt{p}, \mathtt{p} + \mathtt{sizeof}(T) \times n)$ |
| `array_ptr<`$T$`> p: byte_count(`$b$`)` | $[\mathtt{p}, \mathtt{p} + b)$ |
| `array_ptr<`$T$`> p: bounds(`$x, y$`)` | $[x, y)$ |

The interpretation of an *ntarr*'s bounds is similar, but the range can extend further to the right, until a NULL terminator is reached (i.e., the NULL is not within the bounds).

Bounds expressions, like the $n$ in `count(`$n$`)` above, may refer to in-scope variables; `struct` members can refer to adjacent fields in bounds expressions. For soundness, variables used in bounds expressions may neither be modified nor have their address taken, so some legacy idioms may be unsupported. (See Section 3.1 for a worked-out example.)

Checked C also supports polymorphic (generic) types, on both functions and `struct`s. For example, following defines a generic allocation function returning an array of objects of some type $T$.

```
for_any(T) array_ptr<T> alloc(unsigned int s) : byte_count(s);
```

## 2.2 Spatial Safety with Efficiency

Porting an entire program to use checked pointers confers the benefit of *spatial memory safety*, meaning that pointers may not access a buffer outside its designated bounds. The Checked C compiler (implemented as an extension to `clang`) will instrument the program at checked pointer dereferences (load and store) to confirm that (a) the pointer is *not NULL* and (b) that (if an *arr* or *ntarr*) *the dereference is within the range of the declared bounds*. For instance, in the code `if (n>0) a[n-1] = ...` the write is via address $\alpha$ = `a + sizeof(int)` × `(n-1)`. If the bounds of `a` are

`count(u)`, the inserted check will confirm that prior to dereference `a` $\leq \alpha <$ `a + sizeof(int)`$\times$`u`. Failed checks throw an exception.

Oftentimes, inserted checks can be optimized away by LLVM. Consider the above code to be enclosed in another condition, such as, `if (n<u) if (n>0) a[n-1] =`... In such cases, the inserted check can be removed as the outer condition `n<u` already ensures that the dereference is within bounds. The programmer can also use *dynamic bounds casts* `dynamic_bounds_cast`$<T>(e, b)$ to help the optimizer. This code casts $e$ to type $T$ and dynamically checks that either $e$ is NULL, or that the given bounds $b$ are a sub-range of the bounds currently associated with $e$. Effectively the cast asserts a fact which the compiler can leverage statically, but which is soundly verified dynamically. Such casts are especially useful to hoist checks out of loops. The result of all of this is good performance: Experiments on a small benchmark suite [Tarditi et al. 2018] reported average run-time overheads of 8.6% (49.3% in one case); a Checked C port of FreeBSD's UDP and IP stack was found to impose no overhead at all [Duan et al. 2020].

Ruef et al. [2019] and Li et al. [2022] formalized a core model of Checked C and showed what it means for a Checked C program to be spatially safe. In the formalism, all data is represented as an integer annotated with the type the program currently views that it has, e.g., a pointer or a plain number (the annotations are safely erased in the real implementation). The operational semantics premises checked-pointer dereferences on NULL and bounds checks, yielding an exceptional outcome on failure. A spatially safe Checked C program is *sound* in the typical sense: It will either run forever, evaluate to a final value, or it will halt with a NULL or bounds exception; it will never get *stuck*, e.g., by attempting to dereference an integer as if it was a checked pointer. However, spatial safety is partial for partially ported programs, as discussed next.

### 2.3 Backward Compatibility

Checked C's design was inspired by prior safe C dialects such as Deputy [Condit et al. 2007; Zhou et al. 2006] and Cyclone [Jim et al. 2002]. A key departure is that it aims to facilitate *incremental porting*. To this end, Checked C is *backward compatible* with legacy C, which allows checked pointers to be added piecemeal to an existing program, in the style of *gradual typing* [Greenman and Felleisen 2018; Tobin-Hochstadt et al. 2017]. For example, the following is valid Checked C:

```
void foo(int *q) { int x; ptr<int> p = &x; *q = 0; *p = 1; }
```

Spatial safety checks are only added for Checked pointer types, e.g., `p` above.

When a program is not fully ported, the spatial safety guarantee is partial. In particular, a programmer is able to designate regions of code—whole files, single functions, or even single blocks of code—as *checked regions*; these are often designated with a `checked` annotation.[2] Such a region must contain only checked pointer types and adhere to a few other restrictions (e.g., no variadic function calls). The region is sure to be spatially safe in the sense that any run-time safety violation *cannot be blamed* on code in that checked region; rather, the source of the problem was the execution of unchecked code. Ruef et al. [2019] and Li et al. [2022] proved this blame property for their formal model of Checked C. Thus, when an entire Checked C program is in a checked region it is sure to be spatially safe; for partially ported programs, the more code that executes in a checked region, the lower the risk of an exploitable vulnerability.

### 2.4 Interop Types (itypes) and Trusted Casts

Checked C provides *interop types* (aka `itype`s) to allow legacy C functions to be given an *intended* checked type. For example, the Checked C version of `string.h` defines the following prototype:

```
size_t strlen(char *s : itype(nt_array_ptr<char>));
```

[2]You can also designate `unchecked` regions within checked ones, in the style of Rust's `unsafe` blocks.

This type indicates that legacy code may pass to `strlen` a `char *`, while Checked C callers should pass in a `nt_array_ptr<char>` instead.

Itypes can be used on function *definitions* too, not just declarations; e.g., the above prototype could be used with the C code implementing `strlen`. When the function definition appears within a checked region, the function's body is typechecked as if the parameters had the indicated checked types; otherwise, it is typechecked as if it had the unchecked ones. Either way, the compiler will ensure the itype is self-consistent; e.g., itypes like `char *p : itype(ptr<int>)` will be rejected.

The semantics of itypes supports incremental conversion. In particular, if we want to convert module $A$, and it calls into module $B$'s `foo` function, which we don't want to convert just yet, we can annotate `foo` with an itype, and then convert $A$ (and place it in a checked region), which will treat calls to `foo` according to its checked types. At the same time, `foo`'s body will still typecheck without changes. Eventually we will port $B$, including `foo`, whose body we can place in a checked region; then, its itype parameters will be considered as having their checked types. Once all callers of `foo` have been ported, we can swap its itypes for checked ones.

Itypes given to functions outside of a checked region are *trusted*—a spatial safety violation could occur if the function's code does not implement the semantics of the indicated checked type. For example, suppose we give C function `foo` the itype `void foo(int *x: itype(array_ptr<int> count(8))`, but actually `foo` expects `x` to have size 10. Then, callers in checked regions may pass `foo` too-short arrays without complaint. Note that this situation is no different than that of any safe language with a foreign function interface—soundness of safe code is predicated on the foreign code being properly annotated.

Within unchecked regions, programmers can write `assume_bounds_cast`<$T$>$(e, b)$ to cast $e$ to type $T$ with bounds $b$. This has the equivalent compile-time behaviour of the dynamic bounds cast operator but performs no run-time check; as such, it is a potential source of unsoundness. We can think of itypes on functions inside a checked region as inducing an invisible `assume_bounds_cast` on arguments passed in from unchecked callers.

## 3 C TO CHECKED C BY 3C

Our goal is to port a legacy C program to use Checked C. While fully automated conversion might be the ideal, it is impractical. Thus, we have developed an iterative, semi-automated approach.

### 3.1 Porting = Annotation + Refactoring

Porting a program to Checked C involves making two kinds of change: *annotation* and *refactoring*. The first kind leaves the content of the code as is, and involves replacing legacy C types with checked-type alternatives, adding bounds annotations and casts, and labeling (un)checked code regions. For example, consider Listing 1 parts (a) and (c): The former is a legacy C program and the latter is its final Checked C conversion. Note converted function `baz`: its code is unchanged, but its parameters have checked types and its body is labeled `checked`.

While annotations may be all we need, oftentimes we must refactor the code before Checked C will accept it. Consider the code in Listing 2(a). Here, `buf` is a pointer to an array whose (original and updated) size is stored in `*sz`. We might try to annotate this code as shown in Listing 2(b), but Checked C rejects this, disallowing bounds annotations on nested pointers (here, `*buf` is the array). Thus the code must be refactored. There are various ways to do so, but in our experience (Section 6.6) a robust approach is to couple a buffer with its length in a `struct` and adjust the callers accordingly, as shown in Listing 2(c).

While we might hope to fully automate the conversion process, examples like Listing 2 illustrate why doing so is impractical. To automate the refactoring would require soundly inferring the connection of `*buf` to its length `*sz`, abstracting the two into a `struct` and then changing all of the

```
int resize_buf(char **buf, unsigned *sz) {
  unsigned news = round_up(*sz, 64);
  char *newbuf = NULL;
  newbuf = realloc(buf, news);
  *buf = newbuf;
  *sz = news;
  return newbuf != NULL;
}
```

(a) Original code

```
int resize_buf(ptr<array_ptr<char>: count(*sz)> buf,
               ptr<unsigned> sz) {
  unsigned news = round_up(buf->sz, 64);
  array_ptr<char> newbuf: count(news) = NULL;
  ... // as above
}
```

(b) Invalid Annotations

```
typedef struct {
  array_ptr<char> buf: count(sz);
  unsigned sz;
} SIZEBUF;

int resize_buf(ptr<SIZEBUF> buf) checked {
  unsigned news = round_up(buf->sz, 64);
  array_ptr<char> newbuf: count(news) = NULL;
  newbuf = realloc<char>(buf->buf, news);
  buf->buf = newbuf;
  buf->sz = news;
  return newbuf != NULL;
}
// Refactor callers of resize_buf ...
```

(c) Refactored and Annotated Checked C

Listing 2. Porting via refactoring and annotation.

callers to also use it.[3] Doing so might "open a can of worms," precipitating similar refactorings elsewhere; any mistakes will leave a mess for the programmer to clean up or will simply precipitate failure (which is part of the reason that few C refactoring tools exist).

### 3.2 Our approach

Given the practical impossibility of fully automated conversion, we designed a porting process that involves human input. In particular, we developed **3C** to mostly automate the *annotation* portion of porting, and we use it in a way that organizes the needed refactoring. This process may make sense when porting to other gradually typed languages, too. Figure 1 shows the overview of the porting process.

The process has two phases. In the first, we:

(1) Run **3C** on the program, which converts pointers to be checked and adds bounds and other annotations.
(2) Examine the most consequential *root causes* of unconverted pointers; root causes are tabulated by **3C** itself, ordered by influence.
(3) Fix a root cause by refactoring *the original C program*. The fix may be a tweak or a more pervasive change, and may involve adding Checked C annotations.

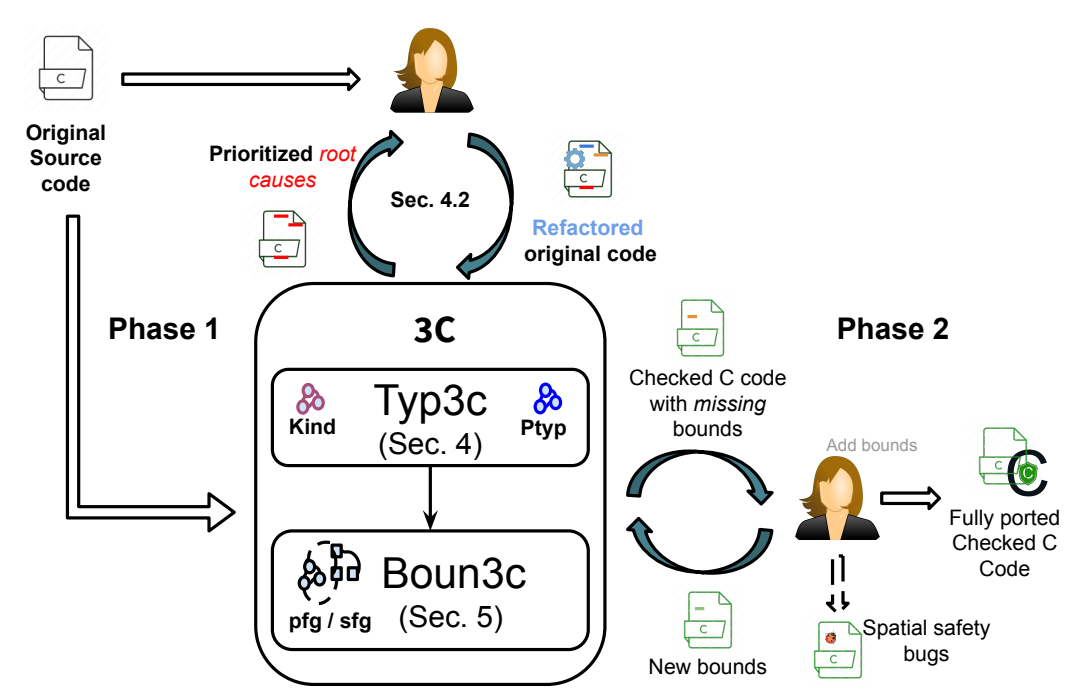

Fig. 1. Overview of the porting process using **3C**.

After each root-cause fix, we iterate the above, rerunning **3C** on the updated C program, which should result in even more converted pointers. Notably, the changed code can always be compiled and tested as usual.

For example, running **3C** on Listing 1(a) produces the program in Listing 1(b). We can see that pointers `g` (line 10) and `p` (line 12) are both unchecked (the latter as part of an itype, for reasons discussed in the next section); the root cause indicated by **3C** is the call `recordptr(p)`. The call effectively casts `p` to type `void *`, which is potentially unsafe, and then `p` is assigned to `g`. We can fix this problem by changing the prototype on line 8 of Listing 1(a) to have the checked type shown on lines 8–9 in Listing 1(c), which indicates that `recordptr` treats its parameter generically (since it is universally quantified), which is safe. Rerunning **3C** on this updated program produces the final and fully converted program shown in Listing 1(c).

[3]CCured and Softbound perform a "refactoring" during compilation that associates each pointer with an added length field/variable; doing so makes `*sz` redundant, avoiding the problem of inferring `*sz` is the intended length. This approach adds overhead, and is not really appropriate when the goal is updating the source program, rather than compiling it.

It will not always be feasible to port the whole program during Phase 1. Porting may take too long, and the programmer may be satisfied with what they have, for now. Thus begins Phase 2. At this point, we have two versions of the program: (a) the original version, which compiles and runs, and due to repeated application of step (3) above will have been refactored and may contain a few Checked C annotations; and (b) the annotated code produced by **3C** when run on this original. Phase 2 involves manually completing the porting process for individual version-(b) files, copying them over to version (a), and testing the result. Checked C's `itypes` (Section 2.4) are leveraged by **3C** to facilitate this process: `itype`-annotated header files are copied over at the start of Phase 2, which means they are compatible with both annotated and unannotated clients. Sometimes it is useful to run **3C** on the copied-over version of a file; this has the effect of propagating manually-introduced changes within it.

The programmer may discover spatial safety bugs during either phase of this process. In particular, she may find that what she thought was a legal bound is rejected by the compiler as invalid. Or, she may find that running the tests triggers a failed run-time check which identifies a spatial safety bug. As discussed in Section 6.6, while porting `tiny-bignum` and `thttpd` we uncovered spatial safety bugs in this manner.

The programmer can stop porting at any time during Phase 2 and will have a runnable, tested, more-safe version of their program.

## 4 TYPE INFERENCE BY TYP3C

**3C** first performs a whole-program analysis called **typ3c** to convert legacy pointers to be checked pointers. It has two parts. The first part determines which pointers *cannot* be converted—we call these *wild*—because they are used in an unsafe way. The second part determines the pointers' type, if checked, i.e., either $ptr$, $arr$, or $ntarr$.

### 4.1 Checked, or Wild

**typ3c** first aims to infer the *kind* of each pointer, which is to say, whether the pointer can be made checked (*chk*) or not (it is *wild*).

*Basic Approach.* Kind inference is essentially a kind of *type qualifier inference* [Foster et al. 2006]. It works by associating each level of a pointer with a qualifier variable $q$ (e.g., `int**` has two levels), and generating a set of constraints $x \sqsubseteq y$ where $x$ and $y$ are either qualifier variables or qualifier literals $chk$ or $wild$. A solution to the constraints is a map from qualifier variables to literals that respects the ordering $chk \sqsubset wild$. We can view the constraints as a *flow graph*: edges $x \longrightarrow y$ correspond to constraints $x \sqsubseteq y$. Variables reachable from $wild$ solve to $wild$; the rest can be $chk$.

Consider the following example.

```
void func(int **y, int *z) {
   z = (int *)5;
   *y = z;
}
```

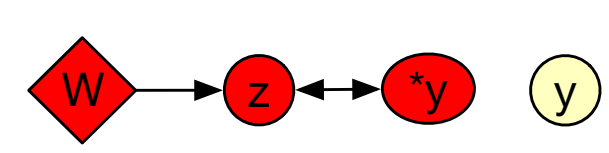

The flow graph, to the right, has four nodes $y$, $*y$, and $z$, for the outer and inner qualifiers of `y` and the qualifier for `z`, respectively, along with the node $W$ (for $wild$). The edge $W \longrightarrow z$ arises from the unsafe use `z = (int *)5` and the bidirectional edge $z \longleftrightarrow *y$ is for the assignment `*y = z`. From this graph, we can determine that $z$, and $*y$ must be $wild$, since they are reachable from $W$, but $y$ can be *chk*, since it is not. Thus `y`'s type in the rewritten program would be `ptr<int *>`; the rest would be unchanged.

*Localizing wildness within functions.* **3C** is designed to automate as much of the annotation process as possible, while isolating and minimizing the work for the human developer to do. To this end, it handles functions and function calls in a novel manner. Consider the following code.

```
int deref(int *y) { return *y; }
int bar(void) { int *p = (int *)5; deref(p); }
```

If we follow the basic approach, the graph will have nodes $W$, $y$, and $p$, and edges $W \longrightarrow p$ (due to assignment `p = (int *)5`), and $p \longleftrightarrow y$ (due to call `deref(p)`). Thus, both `y` and `p` would end up *wild*. Basically, passing *wild* `p` to `deref` has forced its parameter `y` to be *wild* too. On the one hand, doing so seems sensible: passing an unsafe pointer to function can make that function misbehave. On the other hand, the function `deref` is itself completely safe—in the absence of `bar`, if we made `deref`'s parameter have type `ptr<int>` it would typecheck. Thus, making `y` as *wild* would ultimately just make extra work for the programmer, during porting.

**typ3c**'s algorithm determines whether a function parameter is *chk* or *wild* based on that parameter's use by the function. If the *function uses its parameter safely* then **typ3c** gives it a *checked type*. If a caller passes an unchecked pointer, **3C** adds a cast at the call site. We see this in Listing 1(b) with `baz`: This function treats its parameter `c` safely, internally, so it's given checked type `ptr<int>`. However, the call from `bar` on line 26 passes to `c` an unchecked pointer, `g`. Thus, **3C** inserts an `assume_bounds_cast` (Section 2.4) at the callsite. This cast effectively signals to the programmer that there is work to do (and where).

If a *function uses its parameter unsafely*, as with `func` and `z` in our first example, then **typ3c** makes that parameter *wild*. By a similar argument to the one above, we do not want a function's unsafe use of its parameter to force its passed-in arguments *wild* too. Once again, doing so would just make more work for the programmer, who would have to potentially make many manual type changes once `func`'s internals are fixed. To avoid this work, **3C** inserts an `itype`—doing so allows callers to pass in a checked typed argument, despite the internal unsafe use. We see this in Listing 1(b): function `foo` makes parameter `p` *wild* (due to passing it to `recordptr`), and it is given an `itype` so that passed-in arguments can still be *chk*, as with `r` in the call on line 25.

*Implementing localized wildness.* To see how we implement function-localized *wild*ness, consider Figure 2(a), which is the kind flow graph produced when analyzing Listing 1(a). Each node is labeled with the program variable, and each edge is labeled by the line of code that it models.

First, notice that instead of using one node in the graph for each function parameter, we use two. The *internal* node is used when considering code in the function itself, while the *external* node is used at call sites. We can see this in Figure 2(a) for `foo`'s parameter `p`, where the external node is on the left and internal on the right. We connect these nodes with an edge, external to internal.

Second, notice that function calls use a directed edge from the external parameter node to the argument node (assignments within a function use a bidirectional edge as in the basic approach; e.g., see line 17). For `foo`, the call to `baz` on line 18 induces an edge from external `q` to internal `p` (and from external `c` to internal `&m`). This is the reverse of what you'd expect: The flow of data goes from `p` to `q`, but the edge is `q` to `p`. With this arrangement, we can see that there is no path from argument nodes to a parameter's node, nor vice versa. This means that callers and callees are mutually independent, as we want. For example, we can see that despite the fact that `g` is *wild*, and the call `baz(r,g,z)` passes `g` to `baz`'s parameter `c`, the direction of the edge stems the "wildfire." Likewise, even though `foo`'s first parameter `p` ends up *wild*, the call `foo(r,z)` on line 25 does not cause `r` to be *wild*, too.

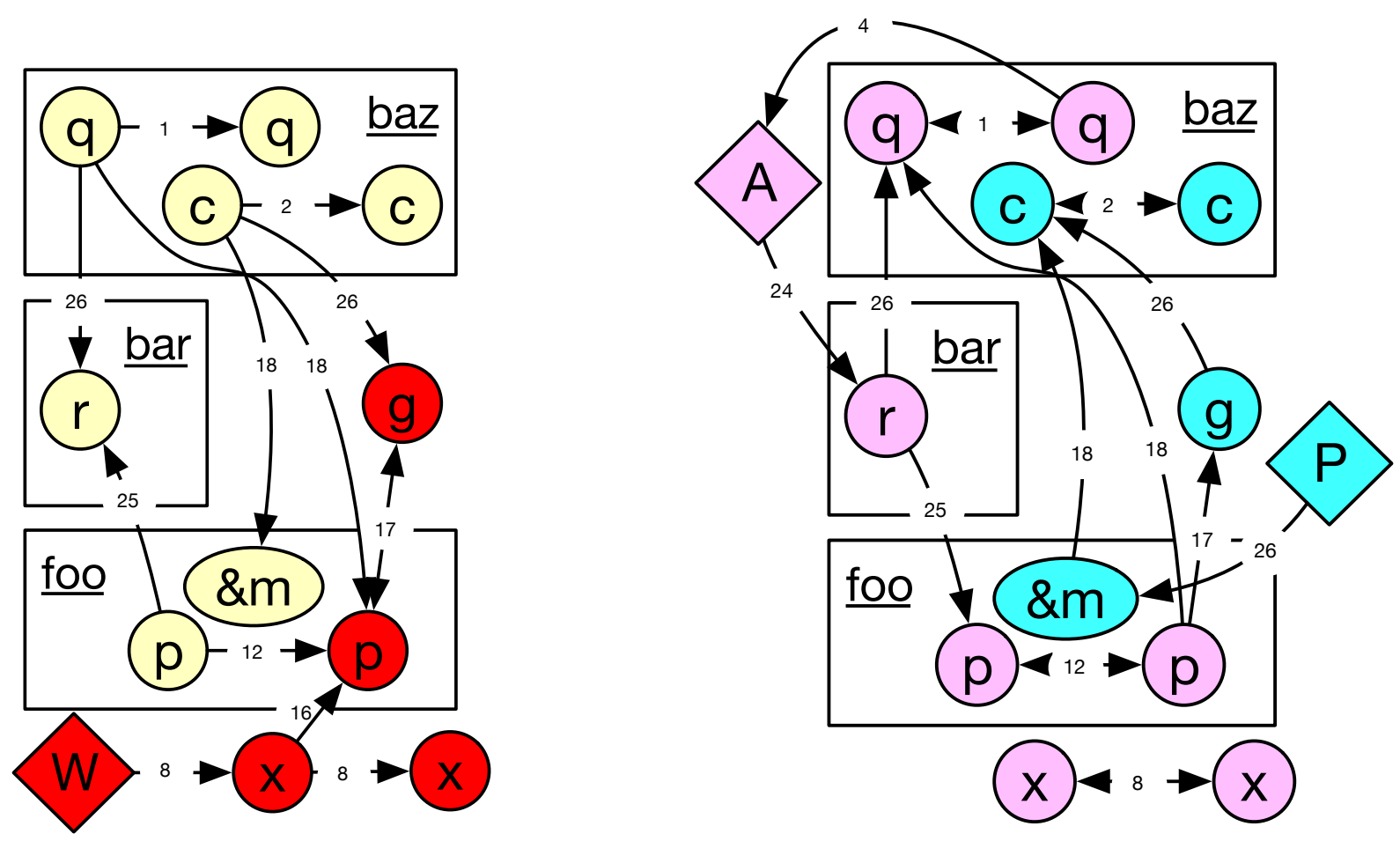

(a) kind graph & solution (red = *wild*, yellow = *chk*)

(b) ptyp graph & solution (violet = *arr*, blue = *ptr*)

Fig. 2. The kind and ptyp graphs and solutions by **typ3c** for Listing 1. Nodes $x$ are variables, except literals $W$ (wild), $A$ (*arr*), and $P$ (*ptr*). An edge $x \xrightarrow{n} y$ represents a qualifier constraint $x \sqsubseteq y$ due to the code at line $n$.

To get the same effect, we might have chosen to generate no edges for a function call, thereby disconnecting arguments and parameters. But the edges are useful when a function's parameter should be *wild with no itype*, and *should* force any passed-in argument to be *wild*. In the example, the `recordptr` function's parameter has a `void *` type, and the function is `extern`. **3C** cannot see (or convert) `recordptr`'s body, so it has no guess as to what checked type or `itype` the parameter should have. Thus **typ3c** adds the edge $W \longrightarrow x_e$ ($x_e$ is the external node of `x`) which because of the reversed call edge will cause arguments to be *wild* (e.g., `p` on line 16). **typ3c** will do the same for the parameters of function *definitions* it finds in system headers (e.g., which are meant to be inlined) or macro definitions, since it cannot rewrite those headers or macros to have checked types.

After solving, **3C** looks at the solutions of the internal and external parameter nodes. When both are *chk*, the parameters gets a checked type; if the external is *chk* but the internal is *wild*, the parameter gets an itype; if both are *wild*, the parameter is left as is. Note that the reverse direction on the call edge—from parameter to argument—means that a parameter could solve to *chk* while its argument solves to *wild*. In this case the **3C** rewriter will insert a cast at the call-site; we see this in Listing 1(b) and Figure 2(a) for the calls to `baz`.

### 4.2 Root Cause Analysis

If **3C** does not convert all pointers to be checked, the developer should strive to refactor their code to convert as many *wild* pointers as possible. To help their work, **typ3c** identifies code that is a *root cause* of *wild*ness, meaning that it is responsible for a direct edge $W \longrightarrow q$ in the graph. This is a place where, for example, an unsafe cast occurs or where an `extern` function's params/return were made *wild*. Fixing a root cause may have positive downstream effects. A pointer can be marked *wild* either directly (e.g., `x` in Figure 2(a)) or indirectly, due to a path from a directly marked pointer (`p` and `g`). Each directly marked *root cause pointer* is identified by **3C** along with the reason it was made *wild*. **3C** will output a report ordered by *influence*, with the root-cause pointers responsible for the most downstream *wild* pointers coming first.

For our example, **3C** will indicate that `recordptr`'s parameter `x` is a root-cause pointer (due to `recordptr` being an `extern` function). Suppose that upon inspecting `recordptr`'s code or man page the developer decides that `recordptr` is safe because it treats its argument generically. Using Checked C's generic types feature, she can manually rewrite the `extern` in the original C to be as shown on lines 8–9 in Listing 1(c). Upon re-running **3C** on the updated program, the edge $W \longrightarrow x$ in the graph will disappear, allowing all pointers to solve to *chk*.

### 4.3 Determining Pointer Type

The second part of **typ3c** determines the *type* of a checked pointer, which is either *ptr*, *arr*, or *ntarr*. It does so using another flow graph, the ptyp graph; the ptyp graph for our example is given in Figure 2(b). Once again, generating and solving the constraints in this graph is basically a novel application of type qualifier inference [Foster et al. 2006].

In this graph, nodes represent pointer qualifiers once again, but now rather than solving to *chk* and *wild*, they will solve to *ptr*, *arr*, *ntarr* (each of which has a representative node, $P$, $A$, and $N$). Parameter nodes are paired again, but they are unified by a bidirectional edge between them. ptyp solutions follow lattice order $ntarr \sqsubset arr \sqsubset ptr$, from least to most general. In the graph, constraints $x \sqsubseteq y$ are written as edges $y \longrightarrow x$. Additional edges arise from code idioms that constrain solutions. For example, line 4 indexes `q` as an array, so that leads to the edge $q \longrightarrow A$, which basically says "`q`'s solution can be at most *arr*;" i.e., it cannot be *ptr*. Line 18 takes the address of `m`; this produces edge $P \longrightarrow \&m$, which basically says that the type of `&m` *must* be *ptr*; it cannot be *arr* or *ntarr*. There is a similar edge lower-bounding to `r` with $A$; this arises from the `malloc` call, whose output could be an array but may not be zeroed out, and thus cannot be trusted as *ntarr*.

A solution $S$ to the ptyp graph is a map from qualifier nodes to *ptr*, *arr*, or *ntarr* such that subtyping constraints indicated by the edges are satisfied. Such a solution can be constructed by a linear-time graph traversal [Rehof and Mogensen 1999]. The *least* solution $S$ is one such that for all alternative solutions $S'$, $S(q) \sqsubseteq S'(q)$, for all $q$; the greatest solution is the reverse. While prior work on qualifier inference mostly focuses on least solutions [Foster et al. 2006, 2002; Shankar et al. 2001] (which is our preference for kind inference), **typ3c** pointer-type inference is different.

*Need for both least and greatest solutions.* Intuitively, we want the greatest solution (or most general type) for checked types. For instance, consider an array dereference: `a[i]`, which introduces the constraint $a \sqsubseteq arr$ in ptyp graph. Given the checked type lattice: $ntarr \sqsubset arr \sqsubset ptr$, the possible values of $a$ are $\{arr, ntarr\}$, i.e., an array or null-terminated array. In the absence of other constraints, we only have the evidence that `a` is an array. Hence, we want to pick *arr*, which is the greatest solution that satisfies $a \sqsubseteq arr$.

However, for return types we want the least solution. For example, consider the function to the right. **typ3c** infers that an array is being allocated in the call to `malloc`, and induces the constraints shown in comments. The greatest solution for *ret* would be *ptr*; the least solution would be *arr*. Choosing *ptr* drops information that could be useful; while *ptr* is correct, it may prevent future uses of `getarr` that need to know that it returns an *arr*. Hence, we want to find the least solution for function return types.

```
int *getarr(int n) {
  int *x = malloc(sizeof(int)*n); // arr ⊑ x
  return x; // x = ret
}
```

*Solving ptyp constraints.* A naive approach of independently picking the least solution for function return types and the greatest for other pointers does not work because of possible interdepedencies.

| | | |
|---|---|---|
| N(x) | = | $x$ |
| N(n) | = | $n$ |
| N($\mathtt{f}_{\mathtt{p}}^{\mathtt{i}}$) | = | $f_p^i$ |
| N(e->f) | = | $f_s$, where type(e) = struct s * |
| N(e.f) | = | $f_s$, where type(e) = struct s |
| ProcExpr($f(e_1, ..., e_n)$) | = | ProcExpr($e_1$), ..., ProcExpr($e_n$), AddSEdge(N($e_1$), N($f_p^1$)), ..., AddSEdge(N($e_n$), N($f_p^n$)) |
| ProcExpr($e_1 = e_2$) | = | ProcExpr($e_1$), ProcExpr($e_2$), AddSEdge(N($e_2$), N($e_1$)) |

(a) Expression processing to create the *sfg*

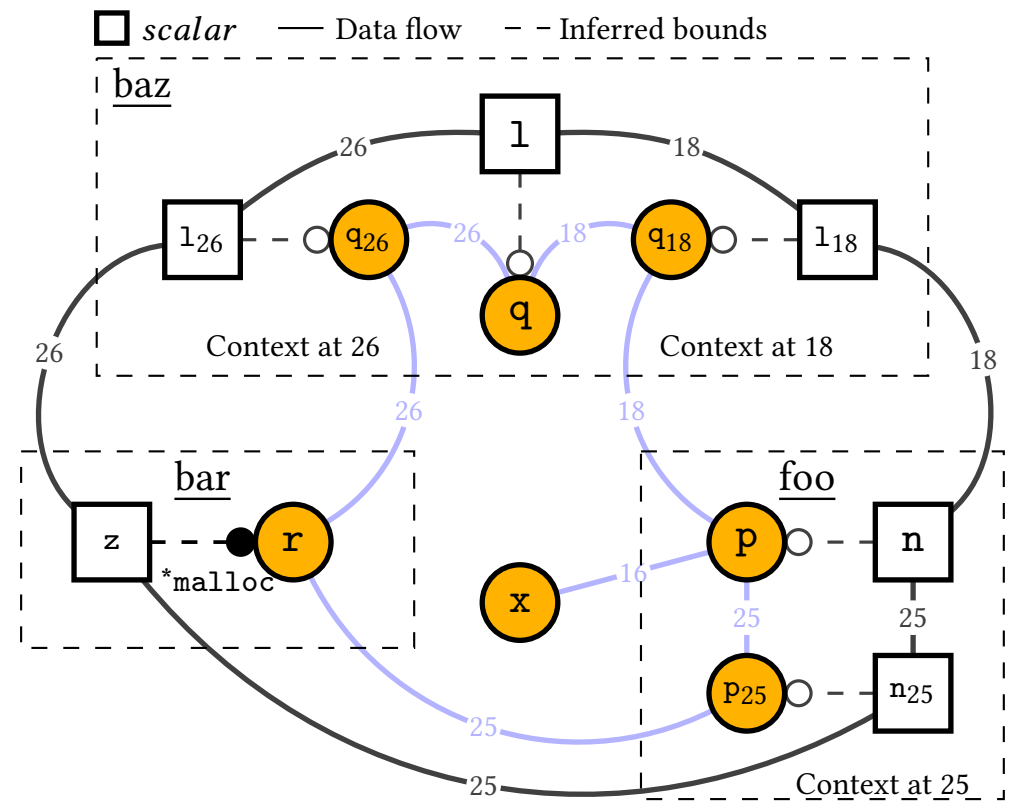

(b) **boun3c**'s *pfg* (circle nodes, only *arr* and *ntarr*) and *sfg* (square nodes) graphs, and solutions (circle-dashed lines)

Fig. 3. Overview of graph creation in **boun3c**.

For instance, a function's return type might depend (through constraints) on one of its parameters' type. We solve ptyp constraints using a novel three-step algorithm. First, the algorithm computes the greatest solution, fixing the solution of function parameters and resetting the others. Then it computes the least solution, fixing it for returns.[4] Finally, it computes the greatest solution for what's left (e.g., local variables and `struct` fields). Solving is linear time for each step.

*Soundness.* **typ3c** aims to be sound, in the sense that its output should be accepted as correct by the Checked C compiler's typechecker. **3C** is not part of the *trusted computing base* (TCB), so mistakes or omissions it makes are not security problems—the Checked C compiler (which is part of the TCB) will confirm that the converted program is correct. **3C**'s soundness argument follows from the observation that checked pointer types are qualified types, and the core qualifier inference algorithm that **typ3c** is based on is sound for the ptyp lattice we use, and works correctly for our amended kind constraints because of the added casts and itypes (Section 2).

## 5 BOUNDS INFERENCE BY BOUN3C

The goal of **boun3c** is to infer a bound for each *arr* and *ntarr* pointer inferred by **typ3c**. The basic idea is to associate each pointer with a possible bound, and then propagate that association consistently. **boun3c** represents a novel kind of *correlation analysis* [Pratikakis et al. 2006].

### 5.1 Generating Flow Graphs

**boun3c** starts by constructing two undirected graphs, the *pointer flow graph pfg* and the *scalar flow graph sfg*, which track the flows of pointer variables and scalar variables (and constants), respectively. The *pfg*'s nodes and edges are isomorphic to those in the ptyp constraint graph, where $x \longrightarrow y$ in the ptyp graph induces an undirected edge $x$ — $y$ in the *pfg*. The *pfg* also contains *context-sensitive call* nodes, one for each argument at a particular call site. We also maintain context-sensitivity for `struct` accesses such that all field accesses using the same base expression (e.g., `e._` and e->_) will be grouped together. We write $\mathbb{P}$ for the set of all *pfg* nodes and $\mathbb{A}$ for the subset of those nodes whose ptyp solution was either *arr* or *ntarr*.

[4]There are a few cases where we do not use least solution; see Appendix A.3.

The *sfg* is like the *pfg* but only considers expressions involving scalar values. Moreover, the *sfg* only considers simple assignments from variables/parameters, fields, or constants to fields, local variables/parameters, and function-returns; it ignores all other expressions, such as `x=y+2` or `func(x+y)`. Figure 3a defines function ProcExpr which processes a function call or assignment to generate *sfg* edges via the AddSEdge($n_1, n_2$) function. The N function translates the allowed expression types into names for nodes in *sfg*. (Note: $\mathtt{f}_\mathtt{p}^\mathtt{i}$ is the *i*th parameter of function $f$ while *n* is an integer constant.) We write $\mathbb{S}$ for the set of all nodes in *sfg*.

*Example.* Figure 3b illustrates *pfg* and *sfg* for the example in Listing 1(a).[5] For ease of understanding, we only show the *arr* and *ntarr* pointers in the *pfg*. The squares and dark solid lines are the nodes and edges of the *sfg*, while the circles and light solid lines are the nodes and edges of the *pfg* and thus require (or are already annotated with) bounds. The dashed lines relate pointers to their possible bounds; these are not part of the initial graph but are rather constructed during inference, as discussed below. We can see that the *pfg* also contains context sensitive call nodes $\mathtt{q}_{18}$, $\mathtt{p}_{25}$, and $\mathtt{q}_{26}$ corresponding to function calls at lines 18, 25, and 26, respectively. The *sfg* has nodes for scalar function parameters and local variables, and also has context-sensitive nodes.

## 5.2 Bounds Inference

**boun3c**'s inference algorithm uses the *pfg* and *sfg* to infer array bounds. To explain it, we must define a few terms first.

*Node scope and visibility.* A pointer v's bounds are defined in terms of variables or scalars within v's lexical scope. We define a pointer's *node scope* $\varpi$ with the function *ns*:

$$ns(\mathsf{v}) = \begin{cases} \text{struct s@c } (\varpi_s^c), & \mathsf{v} \text{ is a field of struct } s \text{ at context } c \\ \text{param of } f\text{@c } (\varpi_p^{f_c}), & \mathsf{v} \text{ is a parameter to function } f \text{ called at context } c \\ \text{local of } f \ (\varpi_l^f), & \mathsf{v} \text{ is local variable of } f \\ \text{global } (\varpi_g), & \mathsf{v} \text{ is a global or constant} \end{cases}$$

The *visibility* (vis) of a scope determines the values that can be used in the bounds declaration of a pointer in the scope. For instance, for a pointer v in function $f$'s local scope ($\varpi_l^f$), v's bounds declaration can contain function locals ($\varpi_l^f$) and global variables or constants ($\varpi_g$); i.e., $\text{vis}(\varpi_l^f) = \{\varpi_l^f, \varpi_g\}$. We use context-sensitive scopes for function parameters and structures. Hence, $\text{vis}(\varpi_s^c) = \{\varpi_s^c, \varpi_g\}$; $\text{vis}(\varpi_p^{f_c}) = \{\varpi_p^{f_c}, \varpi_g\}$; and $\text{vis}(\varpi_g) = \{\varpi_g\}$.

*Bounds map.* The final solution of inference is a map $\beta$ from array pointers to bounds expressions, i.e., $\beta : \mathbb{A} \rightharpoonup b$. A bounds expression $b$ is a pair $(\vartheta, s)$; the second element expresses a numeric value using a node $s \in \mathbb{S}$ while the first expresses the value's units, with $\vartheta \in \{\text{ct}, \text{bt}\}$ indicating either `count` (ct) or `byte_count` (bt).

## 5.3 Algorithm

Pseudocode for bounds inference is given in Algorithm 1. The algorithm operates in several steps. First, SeedBounds (explained next) establishes initial pointer-bounds relationships based on the program text. Next, RunInference iteratively propagates these bounds throughout the program using InferBounds until a fixed point is reached. Heuristics are used to seed bounds for pointers that remain (Section 5.4), and these are (re)propagated.

[5] After correcting the `recordptr` type; see Section 4.2.

$e_1$ = malloc($e_2$);
  if N($e_1$) $\notin \beta \implies \underline{\beta(\text{N}(e_1)) = (\text{bt}, \text{N}(e_2))}$
  if $\beta(\text{N}(e_1)) \neq (\text{bt}, \text{N}(e_2)) \implies \underline{\beta_I = \beta_I \cup \text{N}(e_1)}$
$e_1$ = malloc($e_2$*sizeof($ty$)); $\wedge$ (type($e_1$) = $ty$)
  if N($e_1$) $\notin \beta \implies \underline{\beta(\text{N}(e_1)) = (\text{ct}, \text{N}(e_2))}$
  if $\beta(\text{N}(e_1)) \neq (\text{ct}, \text{N}(e_2)) \implies \underline{\beta_I = \beta_I \cup \text{N}(e_1)}$
TypeName x[n]; $\implies \underline{\beta(\text{N(x)}) = (\text{ct}, \text{N(n)})}$

Seeding bounds at memory allocations

**Algorithm 1: boun3c** bounds inference (Part 1).

**Input:** *sfg*, *pfg*, $\mathbb{A}$
**Output:** $\beta$
1 $\beta, \beta_I \leftarrow$ SeedBounds()
2 $\beta \leftarrow$ RunInference($\beta, \beta_I$, *sfg*, *pfg*, $\mathbb{A}$)
3 $\beta \leftarrow$ TryHeuristics($\beta, \beta_I$)
4 $\beta \leftarrow$ RunInference($\beta, \beta_I$, *sfg*, *pfg*, $\mathbb{A}$)
/* Helper functions. */
5 **Function** RunInference($\beta, \beta_I$, *sfg*, *pfg*, $\mathbb{A}$):
6   $L \leftarrow$ FunctionLocalNodes($\mathbb{A}$)
7   $C_s \leftarrow$ CtxSenNodes($\mathbb{A}$)
8   $P \leftarrow$ CtxSenAndOriginals($\mathbb{A}$)
9   changed $\leftarrow$ true
10   **while** changed **do**
11     $\beta$, $\text{l}_\text{c} \leftarrow$ InferBounds($\beta, \beta_I$, *sfg*, *pfg*, $L$)
12     $\beta$, $\text{cs}_\text{c} \leftarrow$ InferBounds($\beta, \beta_I$, *sfg*, *pfg*, $L \cup C_s$)
13     $\beta$, $\text{p}_\text{c} \leftarrow$ InferBounds($\beta, \beta_I$, *sfg*, *pfg*, $P \cup C_s$)
14     changed $\leftarrow \text{l}_\text{c} \vee \text{cs}_\text{c} \vee \text{p}_\text{c}$
15   **return** $\beta$

*Seed Bounds.* The algorithm initially *seeds* bounds for pointers for which the bounds are directly evident in the program text. One source of such bounds are the itypes on standard library functions. For example, here is `bzero`'s type:

```
itype_for_any(T) void bzero(void *dest : itype(array_ptr<T>) byte_count(n), size_t n);
```

The type indicates that the bounds of the parameter `dest` is a `byte_count` defined by parameter `n`. So on a call `bzero(x,c)`, the algorithm would set the bounds of `x` to (bt,`c`).

The algorithm also seeds bounds for fixed-size arrays and array pointers assigned directly from `malloc`, `calloc`, etc., per the rules in the figure above. Note it uses ct when `sizeof` is used. The algorithm only considers allocator calls with simple expression arguments, as indicated by various N(_) definitions in Figure 3a. Also note that the algorithm discards seed bounds as invalid for array pointers that are assigned from multiple `malloc` calls with (syntactically) distinct sizes, as is the case for `p` in this example.

```
int *p = malloc(sizeof(int)*n); ...
p = malloc(sizeof(int)*x);
```

If discovered, the algorithm stores invalid bounds in the set $\beta_I$ and omits them from further steps.

*Bounds propagation.* Starting with seeded bounds, **boun3c** propagates the bounds in a context-sensitive manner. As shown in Algorithm 1, it first propagates the bounds information within each function between function-local arrays (Line 11). Second, it tries to infer the bounds of context-sensitive nodes from function locals (Line 12). Finally, it propagates bounds information from context-sensitive nodes to the corresponding original nodes (Line 13). The propagation of the bounds is done by the InferBounds function, which aims to compute the set of possible bounds (*CB*) for each of the array pointers. The algorithm for InferBounds function is shown in Algorithm 2. It starts by identifying all array pointers ($A$) that still need bounds (Line 2) and for each pointer $c$, at Line 8, it finds all the neighbor pointers from *pfg* that are identified as arrays (i.e., $\in A$). Next, using reachability in *sfg*, it finds all visible, in-scope nodes (Line 12) to which the bounds of a neighbor flow (Lines 10-13). It maintains the set of common bounds that all neighbors agree on (Line 14). Finally (Lines 18-19), pointers with a single common bound (preferring those in the same scope as the pointer) are updated in the map. We present the algorithms of the other auxiliary functions, i.e., `GetBoundsFlow`, `GetCommonBoundsSet`, and `FindBounds` in Appendix A.4.

Table 1. Examples of **boun3c** heuristics.

| | Consistent Upper Bound | | | Name Prefix | |
|---|---|---|---|---|---|
| Code : | `if (i < 64) { p[i] = ... }` | `if (j >= n) return -1; x = p[j] + 2;` | `for (i=0; i < s; i++) sum += p[i];` | `struct foo { struct bar *p; ... unsigned p_len; }` | `struct baz { int *p; ... unsigned psize; }` |
| $\beta$(p) = | (ct, N(64)) | (ct, N(n)) | (ct, N(s)) | (ct, N(p_len)) | (ct, N(psize)) |

*Example.* For the example in Figure 3b, the only seed bound is `z` for `r` due to the `malloc` call on line 24 of Listing 1(a). It is depicted as a dashed line with a closed circle in the figure. After inserting it in the map, $\beta(r) = (\text{ct}, z)$, the algorithm propagates the bounds to context-sensitive arguments $p_{25}$ and $q_{26}$, which are neighbors to `r` in the *pfg* due to calls on line 25 and 26 respectively. This results in $\beta(p_{25}) = (\text{ct}, n_{25})$ and $\beta(q_{26}) = (\text{ct}, l_{26})$.

Next, the algorithm propagates the bounds from context-sensitive parameter $p_{25}$ to $p$ as $(\text{ct}, n)$. The next iteration propagates the information from $p$ to $q_{18}$. Finally, for $q$, with neighbors $q_{26}$ and $q_{18}$, whose bounds flow into a common variable $l$, the algorithm infers bounds to be $\beta(q) = (\text{ct}, l)$. The computed solution yields the code in Listing 1(c).

**Algorithm 2: boun3c bounds inference (Part 2).**

```
1  Function InferBounds(β, β_I, sfg, pfg, 𝔸_c):
       /* Array pointers that need bounds.                     */
2      A ← {c | (c ∈ 𝔸_c) ∧ (c ∉ β) ∧ (c ∉ β_I)}
       /* Set bounds to a empty set.                           */
3      ∀c ∈ A : CB(c) ← ∅
4      changed ← true
5      while changed do
6          changed ← false
7          for c ∈ A do
8              N_c^a ← pfg.arr_neighbors(c)
9              SSB_c ← ∅
10             for c_i ∈ N_c^a do
11                 B_{c_i} ← FindBounds(c_i, β, CB)
12                 FB_{c_i} ← GetBoundsFlow(c, B_{c_i})
13                 SSB_c ← SSB_c ∪ {FB_{c_i}}
14             B_c ← GetCommonBoundsSet(SSB_c)
15             if B_c ≠ CB(c) then
16                 CB(c) ← B_c
17                 changed ← true
       /* Find the common bounds variable.                     */
18     for c ∈ A do
19         VB_c ← {(ϑ, s) | (ϑ, s) ∈ CB(c) ∧ (ns(s) = ns(c))}
20         if |VB_c| = 1 then
21             β(c) ← first(VB_c)
22         else if |CB(c)| = 1 then
23             β(c) ← first(CB(c))
       /* Flag to indicate new bounds.                         */
24     added ← (∃c | c ∈ 𝔸_c ∧ c ∈ β)
25     return β, added
```

### 5.4 Partial Soundness, Heuristics

The **boun3c** algorithm discussed so far aims to be *partially sound*, in the sense that if a bound is inferred, it is correct; possibly-incorrect bounds are omitted. This argument follows from the soundness argument of *correlation inference*, the core algorithm to which **boun3c** is related [Pratikakis et al. 2011]. Since missing bounds necessitate additional manual work (Section 3), we developed heuristics that, when applied in the presented order, add likely-correct bounds to array pointers *arr* that lack them.

*Consistent upper bound (**CUB**).* The idea is to identify a variable or constant that represents the maximum value of an index used to access a given array. Specifically, if index variables used to access an array $c$ are always upper bounded by the same variable $ub$ then $ub$ will be considered as the ct bound for $c$. Few examples are shown in the left half of Table 1.

*Name prefix (**NPr**).* For an array field of struct $s$ with name $f_n$, we try to find another scalar field $f^2$ with name $f_n^2$ such that $f_n^2$ starts with $f_n$ and contains a count-evoking keyword, e.g., `len` or `size`. This rule can be informally written as: $f_n^2$`.startswith` $(f_n) \wedge f_n^2$`.match("len"|"size")`

Table 2. Benchmark programs. Total Pointers (*TP*) counts *convertible* pointers; e.g., *TP*(`int **p`) = 2.

| Program | Category | Size (SLOC) | Total Ptrs (*TP*) | Num Files (`.c` & `.h`) |
|---|---|---|---|---|
| vsftpd | FTP Server | 14.7K | 1,765 | 78 |
| icecast | Media Server | 18.2K | 2,682 | 72 |
| lua | Interpreter | 19.4K | 4,176 | 57 |
| olden | Data-structure benchmark | 10.2K | 832 | 51 |
| parson | Json parser | 2.5K | 686 | 3 |
| ptrdist | Pointer-use benchmark | 9.3K | 920 | 39 |
| zlib | Compression Library | 21.3K | 647 | 25 |
| libtiff | Image Library | 68.2K | 3,478 | 43 |
| libarchive | Archiving library | 146.8K | 10,269 | 149 |
| thttpd | HTTP Server | 7.6K | 829 | 18 |
| tinybignum | Integer Library | 1.4K | 129 | 7 |
| | **Total** | 319.6K | 26,413 | 542 |

Few examples are shown in the right half of Table 1.

*Next parameter (**NePa**).* Arrays are often passed to functions with their lengths. For an *arr* function parameter `c`, the immediate parameter to its right, say `p`, is considered as `c`'s ct bound if `p` is a scalar, and *not* involved in any arithmetic or bitwise operations (to avoid scalar parameters that are flags).

## 6 EVALUATION

We implemented **3C** as part of the Secure Software Development Project (SSDP)'s open-source fork of Checked C's extended `clang` compiler [repo 2022]. **3C** adds 13K lines of (single-threaded) C++ (per SLOCCount) to the compiler's codebase. Given a set of source files, **3C** runs **typ3c** and **boun3c** and rewrites the files (including `#included` project headers) with Checked C type annotations.

In this section, we evaluate **3C** as follows, measuring

- **(Performance of 3C)** how long it takes to run **3C** on the benchmark programs, and the relative contribution of various phases of **typ3c** and **boun3c** to the total time (Section 6.2);
- **(Effectiveness of typ3c)** how often **typ3c** infers checked pointers, and the contribution of each of our improvements to the result (Section 6.3);
- **(Effectiveness of boun3c)** how often **boun3c** infers bounds on array (*arr*) and null-terminated array (*ntarr*) pointers and how much heuristics help (Section 6.4);
- **(Quality of identified types)** how well **3C** infers annotations compared to those chosen by hand, in previously ported code [Tarditi et al. 2018];

We qualitatively evaluate the **effectiveness of our iterative porting process** by reporting on experience porting `thttpd` (which contains an exploitable vulnerability), `icecast`, and `vsftpd`; as far as we are aware, these are the three largest programs ported to Checked C. We also discuss a port of a smaller program, `tiny-bignum`; the porting process uncovered two spatial safety vulnerabilities therein.

### 6.1 Experimental Setup

We use the programs listed in Table 2 for our evaluation. Most of these programs were suggested by the Checked C team [Microsoft 2019] as good porting targets. All experiments were run on an AMD EPYC 7B12 machine with 8GB RAM running Ubuntu 20.04.

Table 3. Time split (in seconds) of various phases of 3c. We ran each benchmark seven times and picked the median. RC Comp indicates time for computing root causes and SIQR [Bingham 1996] is the Semi Inter-Quantile Range of the total time over seven runs. All percentages are corresponding to the total time.

| Program | Setup | Constraints Building | Constraints Solving | Bounds Inference | Rewriting | RC Comp | Total Time (s) | SIQR (s) |
|---|---|---|---|---|---|---|---|---|
| vsftpd | 1.06 (35.7%) | 0.45 (15.0%) | 0.18 (6.1%) | 0.23 (7.9%) | 0.94 (31.8%) | 0.07 (2.5%) | 2.96 | 0.26 (8.9%) |
| icecast | 6.41 (46.9%) | 1.2 (8.8%) | 1.46 (10.7%) | 1.05 (7.7%) | 2.66 (19.5%) | 0.63 (4.6%) | 13.66 | 1.02 (7.5%) |
| lua | 2.19 (37.7%) | 0.98 (16.9%) | 0.38 (6.6%) | 0.76 (13.0%) | 1.25 (21.5%) | 0.16 (2.8%) | 5.81 | 0.79 (13.5%) |
| olden | 1.57 (63.4%) | 0.23 (9.5%) | 0.22 (9.1%) | 0.1 (4.2%) | 0.25 (10.1%) | 0.06 (2.5%) | 2.47 | 0.04 (1.6%) |
| parson | 0.21 (39.3%) | 0.12 (23.6%) | 0.03 (5.1%) | 0.09 (17.2%) | 0.06 (12.1%) | 0.02 (4.6%) | 0.53 | 0.07 (13.5%) |
| ptrdist | 1.12 (54.2%) | 0.34 (16.3%) | 0.19 (9.3%) | 0.14 (6.6%) | 0.21 (10.4%) | 0.05 (2.5%) | 2.06 | 0.12 (5.7%) |
| zlib | 0.87 (47.3%) | 0.44 (23.9%) | 0.11 (6.2%) | 0.12 (6.5%) | 0.21 (11.7%) | 0.05 (2.5%) | 1.83 | 0.22 (11.9%) |
| libtiff | 3.26 (38.1%) | 1.58 (18.5%) | 0.65 (7.6%) | 0.75 (8.8%) | 1.6 (18.7%) | 0.55 (6.4%) | 8.55 | 1.32 (15.4%) |
| libarchive | 14.63 (33.5%) | 4.02 (9.2%) | 2.91 (6.6%) | 6.2 (14.2%) | 13.94 (31.9%) | 1.31 (3.0%) | 43.74 | 5.75 (13.1%) |
| thttpd | 1.02 (42.2%) | 0.51 (21.1%) | 0.15 (6.1%) | 0.36 (14.7%) | 0.28 (11.6%) | 0.06 (2.6%) | 2.41 | 0.26 (10.6%) |
| tinybignum | 0.24 (52.6%) | 0.07 (14.3%) | 0.06 (12.9%) | 0.02 (4.4%) | 0.06 (12.2%) | 0.01 (3.1%) | 0.46 | 0.05 (10.9%) |
| **Total** | 32.57 (38.5%) | 9.93 (11.8%) | 6.34 (7.5%) | 9.82 (11.6%) | 21.47 (25.4%) | 2.99 (3.5%) | 84.49 | 9.9 (11.7%) |

*Handling Pointers in Macro Expansions.* **3C** rewriting leverages a `clang` library, which unfortunately does not support rewriting within macro definitions or expansions. Thus, **typ3c** labels all pointers $p$ used in macros as *wild* (by adding a $W \longrightarrow p$ constraint). Doing so may induce pointers dependent on $p$ to be *wild*. To sidestep this limitation and faithfully assess **3C** effectiveness, in the next two sections we use a custom tool to expand all uses of macros in the program source (but making no other changes). We emphasize that *macro expansion is not required for 3C*. Running **3C** on the non-preprocessed programs results in a similar detection rate of *chk* pointers, as detailed in Appendix A.2.

### 6.2 Performance

Table 3 shows the time taken (in seconds) by **3C** for each of our benchmarks. The table also shows the split of the total time across various phases. *Setup* comprises parsing, preparing, and typechecking the input files; *Constraints Building* and *Constraints Solving* comprise the two phases of **typ3c**, building and solving the qualifier constraint graph; *Bounds Inference* is **boun3c**'s bounds inference; *Rewriting* comprises rewriting the input files with inferred checked pointers and bounds; and *RC* is the root-cause analysis. We ran **3C** seven times, reporting the median and the Semi Inter-Quantile Range (SIQR) [Bingham 1996] to express timing variation.

For most benchmarks, **3C** took less than 10 seconds with an SIQR $\leq$ 15% of the total time, indicating minor timing variation. These times could be improved, but most are fast enough for interactive porting, since the manual work takes a few minutes or more between runs. Running times generally track a project's code size, but not always. For example, though `icecast` and `lua` have similar SLOC and file counts, their setup times are very different. This is because SLOC counts in Table 2 ignore non project-specific header files, which can be voluminous. We computed project size *post pre-processing* and found that `icecast` is about 900K LOC by this measure while `lua` is 231K; `icecast` relies on a large number of external libraries with many headers included by each of its source files. As another comparison point, `libtiff` has higher SLOC than `icecast`, but its post-processing size is much smaller, at 347K.

### 6.3 Effectiveness of typ3c

As explained in Section 4, **typ3c** generates and solves two sets of constraints: for each pointer $c$, the solution to the kind constraints determines whether $c$ is checked (else it's *wild*), while the solution to the ptyp constraints gives the exact checked type (i.e., *ptr*, *arr*, or *ntarr*).

Table 4. Pointers inferred by **typ3c** to be *chk* (**typ3c**) vs when the node pairs (Section 4.1) on functions are *disabled* (**typ3c**$^f$) vs previous work (CCured) and breakdown of pointer types inferred by **typ3c**.

| Program | Total Pointers (*TP*) | Checked pointers (*chk*) (% of *TP*) typ3c | typ3c$^f$ | CCured | Split of Identified Checked Pointer Types (% of typ3c) ptr | arr | ntarr |
|---|---|---|---|---|---|---|---|
| vsftpd | 1,765 | **1,336 (75.7%)** | 1,226 (69.5%) | 999 (56.6%) | **1,199 (89.7%)** | 44 (3.3%) | 93 (7.0%) |
| icecast | 2,682 | **1,795 (66.9%)** | 1,670 (62.3%) | 1,377 (51.3%) | **1,429 (79.6%)** | 54 (3.0%) | 312 (17.4%) |
| lua | 4,176 | **2,781 (66.6%)** | 2,248 (53.8%) | 1,771 (42.4%) | **2,273 (81.7%)** | 254 (9.1%) | 254 (9.1%) |
| olden | 832 | **721 (86.7%)** | 709 (85.2%) | 709 (85.2%) | **571 (79.2%)** | 130 (18.0%) | 20 (2.8%) |
| parson | 686 | **507 (73.9%)** | 425 (62.0%) | 291 (42.4%) | **405 (79.9%)** | 9 (1.8%) | 93 (18.3%) |
| ptrdist | 920 | **684 (74.3%)** | 652 (70.9%) | 623 (67.7%) | **465 (68.0%)** | 181 (26.5%) | 38 (5.6%) |
| zlib | 647 | **385 (59.5%)** | 375 (58.0%) | 337 (52.1%) | **293 (76.1%)** | 86 (22.3%) | 6 (1.6%) |
| libtiff | 3,478 | **2,111 (60.7%)** | 2,016 (58.0%) | 1,194 (34.3%) | **1,694 (80.2%)** | 177 (8.4%) | 240 (11.4%) |
| libarchive | 10,269 | **6,842 (66.6%)** | 6,190 (60.3%) | 4,924 (48.0%) | **5,532 (80.9%)** | 896 (13.1%) | 414 (6.1%) |
| thttpd | 829 | **634 (76.5%)** | 616 (74.3%) | 449 (54.2%) | **341 (53.8%)** | 57 (9.0%) | 236 (37.2%) |
| tinybignum | 129 | **128 (99.2%)** | 117 (90.7%) | 117 (90.7%) | **110 (85.9%)** | 3 (2.3%) | 15 (11.7%) |
| **Total** | 26,413 | **17,924 (67.9%)** | 16,244 (61.5%) | 12,791 (48.4%) | **14,312 (79.8%)** | 1,891 (10.6%) | 1,721 (9.6%) |

Table 5. Number of required bounds (*RB*) that **boun3c** inferred for *arr* and *ntarr* pointers, and in what phase of inference (*seed*ing, during *flow*, or via *heuristics*).

| Program | Arrays (arr) Require Bounds ($RB_a$) | Inferred Bounds Total (% of $RB_a$) | Technique (% of Total) Seeded | Flow | Heuristics | Null-terminated arrays (ntarr) Require Bounds ($RB_n$) | Inferred Bounds Total (% of $RB_n$) | Technique (% of Total) Seeded | Flow |
|---|---|---|---|---|---|---|---|---|---|
| vsftpd | 30 | 26 (86.7%) | 15 (57.7%) | 6 (23.1%) | 5 (19.2%) | 27 | 18 (66.7%) | 17 (94.4%) | 1 (5.6%) |
| icecast | 29 | 20 (69.0%) | 16 (80.0%) | 4 (20.0%) | 0 (0.0%) | 159 | 59 (37.1%) | 48 (81.4%) | 11 (18.6%) |
| lua | 146 | 79 (54.1%) | 61 (77.2%) | 18 (22.8%) | 0 (0.0%) | 99 | 28 (28.3%) | 18 (64.3%) | 10 (35.7%) |
| olden | 91 | 87 (95.6%) | 68 (78.2%) | 19 (21.8%) | 0 (0.0%) | 0 | 0 (0.0%) | 0 (0.0%) | 0 (0.0%) |
| parson | 2 | 2 (100.0%) | 2 (100.0%) | 0 (0.0%) | 0 (0.0%) | 33 | 22 (66.7%) | 12 (54.5%) | 10 (45.5%) |
| ptrdist | 127 | 91 (71.7%) | 53 (58.2%) | 38 (41.8%) | 0 (0.0%) | 11 | 7 (63.6%) | 4 (57.1%) | 3 (42.9%) |
| zlib | 52 | 50 (96.2%) | 37 (74.0%) | 12 (24.0%) | 1 (2.0%) | 1 | 0 (0.0%) | 0 (0.0%) | 0 (0.0%) |
| libtiff | 65 | 62 (95.4%) | 42 (67.7%) | 20 (32.3%) | 0 (0.0%) | 145 | 145 (100.0%) | 144 (99.3%) | 1 (0.7%) |
| libarchive | 449 | 347 (77.3%) | 255 (73.5%) | 83 (23.9%) | 9 (2.6%) | 112 | 40 (35.7%) | 27 (67.5%) | 13 (32.5%) |
| thttpd | 31 | 26 (83.9%) | 19 (73.1%) | 7 (26.9%) | 0 (0.0%) | 127 | 96 (75.6%) | 61 (63.5%) | 35 (36.5%) |
| tinybignum | 2 | 2 (100.0%) | 2 (100.0%) | 0 (0.0%) | 0 (0.0%) | 13 | 13 (100.0%) | 13 (100.0%) | 0 (0.0%) |
| **Total** | 1,024 | **792 (77.3%)** | **570 (72.0%)** | **207 (26.1%)** | 15 (1.9%) | 727 | **428 (58.9%)** | **344 (47.3%)** | **84 (11.6%)** |

*6.3.1 kind constraints.* The left half of Table 4 shows that **typ3c** was able to infer 67.9% of all pointers as *chk*, which is 19.5% more than the unification-based algorithms used in past work, e.g., **CCured** [Necula et al. 2005]. We attribute the improved detection rate to our two improvements over the default approach. Specifically, as indicated by the **typ3c**$^f$ column, we were able to infer an additional 6.4% (67.9%-61.5%) of *chk* pointers by maintaining two constraint variables for parameter and return values (Section 4.1), and an additional 13.1% (61.5%-48.4%) of *chk* pointers over unification-style analysis (the "basic approach" of Section 4.1).

Although **typ3c** was able to infer more checked pointers than other techniques, there are still a considerable number (32.1%) of pointers left as *wild*. These arise because of relatively fewer *root-cause pointers*—only 2,333 (8.8%), which can be the focus of an iterative port (Section 4.2). Many of these root causes arise for the same reasons, such as unsafe casts, and conversion to `void`; there were 91 unique reasons in total. Appendix A.1 presents a comprehensive analysis of these *wild* pointers.

*6.3.2 ptyp constraints.* The right half of Table 4 shows the pointer types–*ptr*, *arr*, or *ntarr*—that **typ3c** infers for *chk* pointers. The majority (79.8%) are *ptr*, with *arr* and *ntarr* roughly equal at nearly 10%. These types are determined by **typ3c**'s three-step solving algorithm, described in Section 4.3.

As explained in Section 4.3, **typ3c**'s solving algorithm produces the most general types for all checked pointers. Had it used the least solution instead (as is typical in qualifier inference [Foster et al. 2006]), the solution would have been very different: 37% *ptr*, 37% *arr*, and 26% *ntarr*. This solution is still valid, but the increased number of *arr* and *ntarr* pointers has at least three downsides. First, when these are used for function parameter types, they limit future callers; e.g., a function `foo(ptr<int> p)` is more general (can be called with more pointer types) than `foo(nt_arr_ptr<int> p)`. Second, *ntarr* occurs relatively rarely in C code, in our experience; hence, even when returning an *ntarr* is strictly more general than returning a *ptr*, it was probably not what the programmer intended. Last, *ntarr* pointers require bounds; if the programmer's intention was really for most of these to be *ptr*, then **boun3c** is not going to succeed. Further discussion on these points can be found in Appendix A.3.

### 6.4 Effectiveness of boun3c

Table 5 tabulates the bounds annotations inferred by **boun3c** for *arr* and *ntarr* pointers inferred by **typ3c**. We categorize inferred bounds according to whether they were initial *seeds* (Section 5.3a), determined by propagating *flows* (Section 5.3b), or chosen by *heuristics* (Section 5.4). We limit our attention to pointers that *require bounds* (*RB*). In Checked C, bounds attach to the outermost type, and nested *arr* / *ntarr* pointers cannot have them; for example,

```
array_ptr<nt_array_ptr<char>> argv : count(argc)
```

has no bound for the inner *ntarr*.

*Array pointers (arr).* On average, **boun3c** successfully identified bounds for 77.3% of the *arr* pointers.

Table 6. Effectiveness of heuristics in bounds inference.

| Program | Total | NPr | NePa | CUB |
|---|---|---|---|---|
| vsftpd | 5 | 0 (0.0%) | 0 (0.0%) | 5 (100.0%) |
| zlib | 1 | 0 (0.0%) | 1 (100.0%) | 0 (0.0%) |
| libarchive | 9 | 2 (22.2%) | 3 (33.3%) | 4 (44.4%) |
| **Total** | 15 | **2 (13.3%)** | **4 (26.7%)** | **9 (60.0%)** |

Only 1.89% of inferred bounds owed to using heuristics for three benchmarks. Table 6 shows the relative effectiveness of each of our heuristics for the affected benchmarks. We manually checked a sample of these and confirmed them to be valid. An example is shown in Listing 3, **boun3c**'s *consistent upper bound* heuristic noticed that the index of the `p_in` *arr* is always upper bounded by another parameter `in_len` in this `vsftpd` code (Listing 3 (a)). The inferred bounds on `p_in` are shown in Listing 3 (b).

Indeed, **boun3c** performed well on average, but less well on `icecast` (69% of bounds inferred) and `lua` (54.1%). The situation was similar in both cases: most arrays are allocated using an arithmetic expression, as in this example:

```
ptr = (char *)malloc(n*m);
```

The bounds of `ptr` should be `n*m`, but **boun3c** can only infer bounds that are a single variable or constant. This missing seed bound likely had negative downstream effects.

```
struct bin_to_ascii_ret
vsf_ascii_bin_to_ascii(
 const char *p_in,
 ...,
 unsigned int in_len,...) {
  ➔while (indexx < in_len) {
   char the_char = p_in[indexx];
   ...
  }
}
```

(b) Code matching (➔) **CUB** heuristic

```
struct bin_to_ascii_ret
vsf_ascii_bin_to_ascii(
 (✔CUB)_Array_ptr<const char> p_in : count(in_len),
 ...,
 unsigned int in_len,...)
```

(c) Bounds inferred by **CUB**

Listing 3. Bounds inferred by Consistent Upper Bound (**CUB**) heuristic in `vsftpd`.

*Null-terminated array pointers (ntarr).* **boun3c** inferred bounds for 58.9% of the *ntarr* pointers, on average. Although the detection is relatively less than regular *arr* pointers, we do not consider this as necessarily inferior. Most of the *ntarr*s in our benchmark programs are strings whose bound is discovered through use, so no explicit bound is needed; e.g., functions such as `argparse`, which process a string byte by byte and checking for the null terminator as they go.

This is good because maintaining explicit length variables increases the burden on the developer to update the length on every string manipulation and consequently increases the chances of introducing bugs. One of the effective ways to infer the length of `char*` *ntarr*s could be based on specific handling of `str*` library functions. Doing so has limited ceiling at present, though, due to weaknesses in Checked C's ability to reason about length-extending arithmetic (which the compiler team is addressing).

*Inferring other types of bounds annotations.* **boun3c** is currently limited to inferring `count()` annotations on (nt)arrays; unfortunately, if `p` is modified by pointer arithmetic (e.g., `p++`), it cannot be given such a bounds annotation. We have been developing a feature that infers `bounds()` annotations, for pointers used in arithmetic. It works by first attempting to find another pointer in the same scope that could act as a lower bound. For example, if `p` might have had bound `count(c)` but could not because `p` is subject to pointer arithmetic, but pointer `q` is a lower bound for `p` then `p`'s bound can be `bounds(q,q+c)`. If no such `q` exists, we can replace the introduction of `p` within a function with a fresh `q` (having a `count(c)` bound), and then introduce `p` in the same scope as a local variable that uses `q` as its lower bound, i.e., `array_ptr<T> p : bounds(q,q+c) = q`. Then pointer arithmetic on `p` in the sequel will be accepted. We ran a prototype version of this feature over our benchmarks and found that it adds 134 additional bounds, increasing the total percentage of inferred bounds by 17%, over all the benchmarks. We plan to develop this feature further in future work.

### 6.5 Understanding Annotation Quality

As mentioned in Section 3, completely porting a program from C to Checked C involves both *refactoring* and *annotation*; **3C** aims to automate the latter. Its effectiveness at doing so can be negatively impacted by the presence of certain idioms in the original source code. For example, the use of a custom allocator will harm **typ3c** performance because `malloc` has a generic itype but the custom allocator will not, and it will harm bounds inference because **boun3c** will not see a seed bound (Section 5.3).

As such, an unchanged C program constitutes the worst case for **3C** performance, whereas a fully refactored (but unannotated) program constitutes the best case. To understand **3C**'s performance in these two situations, we carried out an experiment using Ptrdist and Olden benchmarks, which were previously ported to Checked C [Tarditi et al. 2018]. Table 7 contains results; we explain the experiment as we explain the table.[6]

Table 7. Comparison of the effectiveness of **3C** against *complete* manual port of Olden and Ptrdist benchmarks. We report the effort in terms of number of *Refactored* source lines, number of lines *Annotated* with Checked C pointer types along with number of checked *pointers* annotated, added *Bounds* and *Casts*. Refer to Section 6.5 for the meaning of different *Variation*s.

| Program | Variation | Source changes | | | Pointers | | | | Bounds | Casts |
|---|---|---|---|---|---|---|---|---|---|---|
| | | Refactored | Annotated | Left | ptr | arr | ntarr | wild | | |
| **Olden** | | | | | | | | | | |
| bh | manual | 136 (10.44 %) | 45 (3.44 %) | - | 135 | 54 | 3 | 0 | 49 | 10 |
| bh | 3c (revert) | - | 48 (3.66 %) | 30 (2.29 %) | 135 | 54 | 3 | 0 | 49 | 11 |
| bh | 3c (orig) | - | 103 (7.90 %) | 105 (8.03 %) | 132 | 54 | 3 | 0 | 49 | 16 |
| bisort | manual | 57 (21.43 %) | 35 (13.73 %) | - | 44 | 2 | 2 | 0 | 0 | 0 |
| bisort | 3c (revert) | - | 34 (13.33 %) | 1 (0.39 %) | 44 | 2 | 2 | 0 | 0 | 1 |
| bisort | 3c (orig) | - | 51 (19.17 %) | 49 (19.22 %) | 32 | 2 | 2 | 1 | 0 | 3 |
| em3d | manual | 170 (37.04 %) | 88 (17.74 %) | - | 61 | 31 | 3 | 0 | 24 | 6 |
| em3d | 3c (revert) | - | 88 (17.74 %) | 34 (6.88 %) | 64 | 29 | 2 | 0 | 20 | 8 |
| em3d | 3c (orig) | - | 76 (16.56 %) | 158 (34.50 %) | 63 | 30 | 2 | 2 | 17 | 12 |
| health | manual | 42 (11.38 %) | 57 (15.62 %) | - | 70 | 7 | 3 | 0 | 4 | 0 |
| health | 3c (revert) | - | 57 (15.62 %) | **0 (0.00 %)** | 72 | 6 | 2 | 0 | 4 | 0 |
| health | 3c (orig) | - | 57 (15.45 %) | 24 (6.50 %) | 72 | 6 | 2 | 0 | 4 | 3 |
| mst | manual | 133 (40.43 %) | 28 (8.33 %) | - | 44 | 8 | 4 | 0 | 6 | 3 |
| mst | 3c (revert) | - | 28 (8.33 %) | 4 (1.19 %) | 46 | 7 | 3 | 0 | 5 | 1 |
| mst | 3c (tweak) | 19 (5.78 %) | 52 (16.51 %) | 91 (28.53 %) | 39 | 13 | 2 | 5 | 9 | 5 |
| mst | 3c (orig) | - | 42 (12.77 %) | 116 (34.94 %) | 29 | 14 | 2 | 16 | 8 | 3 |
| perimeter | manual | 34 (10.27 %) | 10 (3.03 %) | - | 26 | 2 | 2 | 0 | 0 | 0 |
| perimeter | 3c (revert) | - | 10 (3.03 %) | **0 (0.00 %)** | 26 | 2 | 2 | 0 | 0 | 0 |
| perimeter | 3c (orig) | - | 23 (6.95 %) | 26 (7.81 %) | 26 | 2 | 2 | 1 | 0 | 1 |
| power | manual | 53 (11.67 %) | 30 (6.55 %) | - | 35 | 21 | 1 | 0 | 8 | 0 |
| power | 3c (revert) | - | 30 (6.55 %) | **0 (0.00 %)** | 41 | 15 | 1 | 0 | 8 | 0 |
| power | 3c (orig) | - | 55 (12.11 %) | 37 (8.08 %) | 41 | 15 | 1 | 0 | 8 | 4 |
| treeadd | manual | 30 (20.98 %) | 16 (11.03 %) | - | 12 | 2 | 2 | 0 | 0 | 0 |
| treeadd | 3c (revert) | - | 16 (11.03 %) | **0 (0.00 %)** | 12 | 2 | 2 | 0 | 0 | 0 |
| treeadd | 3c (orig) | - | 18 (12.59 %) | 24 (16.67 %) | 12 | 2 | 2 | 2 | 0 | 3 |
| tsp | manual | 68 (16.15 %) | 10 (2.38 %) | - | 66 | 3 | 3 | 0 | 0 | 0 |
| tsp | 3c (revert) | - | 9 (2.14 %) | 1 (0.24 %) | 68 | 2 | 2 | 0 | 0 | 0 |
| tsp | 3c (orig) | - | 42 (9.98 %) | 31 (7.35 %) | 68 | 2 | 2 | 1 | 0 | 1 |
| **Ptrdist** | | | | | | | | | | |
| anagram | manual | 90 (26.87 %) | 52 (14.48 %) | - | 16 | 32 | 4 | 4 | 21 | 0 |
| anagram | 3c (revert) | - | 28 (7.80 %) | 37 (10.36 %) | 15 | 12 | 7 | 22 | 6 | 6 |
| anagram | 3c (orig) | - | 28 (8.36 %) | 97 (28.96 %) | 12 | 13 | 7 | 12 | 6 | 3 |
| ft | manual | 147 (16.63 %) | 122 (13.74 %) | - | 169 | 2 | 1 | 0 | 1 | 4 |
| ft | 3c (revert) | - | 122 (13.74 %) | **0 (0.00 %)** | 169 | 2 | 1 | 0 | 1 | 0 |
| ft | 3c (orig) | - | 126 (14.25 %) | 146 (16.57 %) | 169 | 2 | 1 | 0 | 1 | 4 |
| ks | manual | 85 (17.03 %) | 35 (7.06 %) | - | 56 | 15 | 6 | 0 | 13 | 4 |
| ks | 3c (revert) | - | 35 (7.06 %) | 11 (2.22 %) | 57 | 15 | 4 | 1 | 13 | 3 |
| ks | 3c (orig) | - | 64 (12.83 %) | 86 (17.23 %) | 56 | 15 | 5 | 1 | 13 | 8 |
| yacr2 | manual | 280 (12.69 %) | 157 (7.07 %) | - | 15 | 135 | 5 | 0 | 57 | 0 |
| yacr2 | 3c (revert) | - | 195 (8.78 %) | 143 (6.44 %) | 54 | 93 | 4 | 4 | 19 | 42 |
| yacr2 | 3c (orig) | - | 190 (8.61 %) | 389 (17.64 %) | 53 | 88 | 4 | 13 | 10 | 62 |

First, we started with the original program, and the manual port. This is the first row in each program grouping in the table. The rightmost columns indicate how many *ptr*, *arr*, etc. pointers, bounds, and other Checked C features are in the Checked C version. The Source Changes *Refactored* column indicates the number of lines that changed from the original program, ignoring annotations. To compute this number, we *reverted* the manual port from Checked C back to C, stripping away the Checked C annotations, and counted the updated/added/deleted lines (ignoring whitespace and linebreaks) in the diff with the original. The *Annotated* column considers the diff between the reverted version and the manual port. Looking at bh, 136 lines were refactored, and 45 were annotated. These 45 annotated lines involved 192 pointers, per the rightmost columns of the table; this is because multiple pointers occur on the same line, or share the same (checked) typedef.

[6]These results are over macro-expanded code; without macro expansion the diffs tend to be smaller. We also do not count one-off `unchecked` annotations in any of the diffs, since **3C** does not infer these.

The latter 3 lines of a grouping capture the performance of **3C**. The *3c(revert)* line considers running **3C** on the reverted version; as mentioned above, this is the best-case scenario. *Annotated* indicates how many lines **3C** annotated, and the pointer counts indicate annotation inference counts. The *Left* column counts the diff between the **3C**-annotated version and the manual port. A diff of 0 in this column indicates that **3C** perfectly inferred (highlighted) all of the pointers and bounds in the manual port; we see this result for health, perimeter, power, treeadd, and ft, while bisort, mst, and tsp had 1 or 4 diffed lines, only. anagram does the worst, with 10.36% of the lines to change, left; it uses a complicated pointer arithmetic scheme that requires Checked C bounds expressions that **boun3c** is unable to infer.

The *3c(orig)* line considers **3C** when run on the original program, prior to refactoring it; this matches the scenario in which we ran **3C** in Sections 6.3 and 6.4, above, and is its worst-case scenario. We can see that *3c(orig)* tends to leave more *Left* compared to *3c(revert)*, in part because the refactoring from *manual* is still to be done. Such refactoring sometimes involves adding new local-variable pointers or changing parameter types, e.g., to support bounds for pointer arithmetic; this is what happens with anagram, and is the reason that there are more pointers overall, and more wild pointers in particular, in the *3c(revert)* row than the *3c(orig)*. The lack of refactoring also tends to make pointer and bounds inference a little worse. Nevertheless, the results for *3c(orig)* are in the same ballpark as *3c(revert)*, providing some context for the results given in Sections 6.3 and 6.4.

Lastly, for mst, we introduce a *3c(tweak)* case, illustrating the benefits of our preferred porting process's Phase 1 (Section 3): After running **3C** on the original code, the developer changes that original code according to an influential root cause, and then reruns **3C**. Here, we switched to use the system malloc rather than the custom allocator that was added; the size of the change is in the *Refactored* column. This change permitted a substantial number of additional checked pointers to be inferred, and an additional bound.

The rightmost two columns tabulate added bounds and casts. The general trend is that the *manual* port tends to have the most bounds, the *3c(revert)* version has equal or fewer, and the *3c(orig)* version has fewer still. One interesting exception is mst. This is because **3C** 's handling of typedefs involving arrays allows it to infer more bounds than were present in the manual port; the programmer had left off bounds entirely and worked around doing so by adding unchecked blocks. **3C** discovered an improvement to this approach.

### 6.6 Porting Programs with 3C

We used **3C** to help carry out iterative, complete ports of four programs: vsftpd, thttpd, icecast, and tiny-bignum. For the vsftpd, thttpd, and icecast we followed the two-phase workflow discussed in Section 3; tiny-bignum converted almost entirely in one step. Table 8 summarizes the impact of each phase (explained below). All four ports, including the revision history detailing each step, are freely available.[7]

These ports were carried out by the second (vsftpd and thttpd), fourth (icecast), and third (tiny-bignum) authors. The first two of these, who carried out the bulk of the work, are recent computer science degree holders, and are very proficient in C and Linux. The third is more senior, but tiny-bignum ended up requiring almost no manual effort. All three authors were familiar with

[7]https://github.com/secure-sw-dev/{checked-vsftpd,checkedc-icecast,checkedc-thttpd,checkedc-tiny-bignum-c}

Table 8. Impact of various phases of **3C**

| Program | Step | Source Changes (LOC) | | Pointers | | | | Root Causes | |
|---|---|---|---|---|---|---|---|---|---|
| | | Manual | 3C | ptr | arr | ntarr | wild | Num | Avg |
| vsftpd | 3c (Initial) | - | 1760 | 1220 | 46 | 98 | 441 | 304 | 21.4 |
| | Phase 1 | 367 | 1616 | 1261 | 82 | 179 | 290 | 224 | 4.5 |
| | Phase 2 | 1889 | | 1407 | 177 | 240 | 97 | 96 | 1.2 |
| thttpd | 3c (Initial) | - | 704 | 338 | 57 | 236 | 198 | 136 | 29.6 |
| | Phase 1 | 708 | 771 | 392 | 75 | 348 | 58 | 53 | 1.9 |
| | Phase 2 | 1450 | | 398 | 87 | 468 | 15 | 25 | 1.6 |
| icecast | 3c (Initial) | - | 2102 | 1424 | 54 | 312 | 887 | 1142 | 51.2 |
| | Phase 1 | 168 | 5529 | 1667 | 62 | 330 | 636 | 874 | 37.0 |
| | Phase 2 | 2592 | | 1829 | 70 | 523 | 328 | 266 | 4.0 |

**3C** and contributed to its development, but none were familiar with the target programs before trying to port them. We view our experience as showing promise; future controlled studies could provide more definitive results. We do have some informal experience with non-authors using **3C** on other programs, and that experience affirmed our approach.

*vsftpd.* The first row of the table shows the effect of initially running **3C** on `vsftpd`: It inferred 1220 *ptr*, 46 *arr*, and 98 *ntarr* checked pointers, compared to 441 *wild* pointers,[8] with 304 identified root causes of wildness influencing, on average, 21.4 other pointers to be deemed wild. 1760 LoC were updated by **3C**.

During Phase 1, we examined the most influential root causes and addressed them via refactorings or annotations to the *original code*, ultimately changing 367 lines, with 1616 more updated by **3C**.[9] As shown in the second `vsftpd` row of the table, doing so increased the total count of checked pointers and reduced the influence of the remaining root causes.

A common root cause was casts to/from `void` pointers, especially when using "generic" C functions like `malloc` returning a `void *`. While the Checked C library headers use an `itype` to treat `malloc`'s return as generic (e.g., see the `recordptr` prototype in Figure 1(c)), `vsftpd` wraps most standard library functions to include defensive checks, and these wrappers lack the annotation. Some of the simplest generic types are inferred automatically by **3C**, but some must be added manually. Other `void` pointers have more subtle constraints on how they can be instantiated which cannot be expressed in Checked C. For instance, `vsftpd` uses `void` pointers for code that can operate generically on either ipv4 or ipv6 addresses, which differ in size. Checked C generics only work on pointed-to values (e.g., $\forall T$, `array_ptr<T>`) not on values themselves, so we needed to resort to a non-checked idiom. In these cases, however, it is still possible to ensure spatial memory safety, if not type safety, by using the checked `array_ptr<void>` type. This type, when accompanied by an appropriate bounds expression will protect against out-of-bounds memory accesses.

In Phase 2, we completed the port starting from the **3C**-converted code and proceeding one file at a time. The final `vsftpd` row of Table 8 shows that this took 1889 lines of mostly manual changes. Most of this work involved adding missing bounds that **3C** could not infer. For example, `vsftpd` provides a string library that uses a `struct` to represent a dynamically resizable string. This `struct` contains both the capacity and the current length of the string; **3C** is not able to infer which is the proper bound. We also sometimes needed to introduce pointers to act as the bounds in `bounds(...)` annotations, and do other small refactorings.

[8] These counts, and others in the table, slightly differ from the counts in Table 4 because there we expanded macros first, to show **3C**'s full capabilities. Also note that the total number of pointers can change across rows in Table 8 because pointer-containing code is added/deleted during a port.

[9] The latter count, for `vsftpd` and the other programs, includes `vsftpd`-internal header files that **3C** automatically updated with `itypes` just prior to the start of Phase 2, thereby supporting both checked and unchecked (i.e., not yet ported) clients.

At the end of the port, 97 unchecked pointers remained for which a Checked C idiom did not apply; these were placed in `unchecked` blocks. In some cases, these blocks included calls to potentially unsafe external library functions. Some calls could be made safe by defining an `itype` for these functions, but for others no such itype is possible, e.g., to functions involving variable length arguments. We also annotated blocks `unchecked` when they contained trusted casts. For example, a bounds cast is required to expand the bounds on a string to be equal to the length of the string determined by `strlen`—the Checked C compiler is not yet smart enough to figure this out. Sometimes casts are needed to/from `void *` arguments, e.g., when implementing a kind of existential type for functions passed `fork` or signal handlers. In all cases, we scrutinized these blocks carefully to convince ourselves they were safe.

*thttpd.* `thttpd` required more pervasive phase-one changes than `vsftpd` because of its use of risky string manipulation code. The two most prominent root causes were uses of the unsafe string library functions `strcpy` and `strcat`, which affected 81 and 73 pointers respectively. Another frequent root cause was the use of the `&` operator on strings with bounds, to be able to dynamically resize them; unfortunately, this is an idiom that Checked C disallows (see the example at the end of Section 2.1). We decided to refactor the string management code to something safer, taking inspiration from `vsftpd`'s safe string library (see Figure 2(c)). Smaller fixes were made, too, including adding itypes for standard library functions. During Phase 1 we changed 708 lines, reducing the number of unchecked pointers from 198 to 58, with 771 additional changes introduced by **3C**.

During Phase 2, our main task was to complete bounds annotations needed, and to make further adjustments (or add unchecked blocks) to the refactored string manipulation code, which often involved pointer arithmetic. Generally speaking, **3C** and Checked C could stand to improve their handling of such code. **3C** fails to leverage *manual* string manipulation idioms; e.g., no constraints are generated based on expressions comparing or assigning an index in the array to null (`(p + 1) == '\0'`; or `p[i] = '\0'`). In cases where **3C** correctly infers *ntarr* types, Checked C is limited by the code patterns it can accept, often because information about a string's length is not apparent to it. As already mentioned, it does not relate calls of `strlen` to a string's bound, and it cannot express that functions such as `strchr` and `strstr` return a pointer that is an address within the bounds of the parameter string. Planned compiler improvements will address some of these issues [Li et al. 2022].

An interesting aspect of converting `thttpd` is that the version we ported to Checked C contained the known CVE-2003-0899 for a spatial safety violation. The extension of **3C** (Section 6.4) that can infer simple `bounds(p,q)`-style bounds annotations inferred a proper bound for the buffer manipulated by the buggy code. As such, an overrun would have been prevented by a run-time check. Even without this feature, a hand-added bound during Phase 2 would block exploitation.

*icecast.* `icecast` makes extensive use of third-party libraries, including `libxml`, `libogg`, and `libvorbis`. Since Checked C does not provide `itype`-annotated header files for these libraries, the initial run of **3C** yielded a significant number of *wild* pointers for uses of them (pointed out as root causes). We used **3C** to automatically construct `itype` annotations for these libraries' headers by copying the headers into a local directory in the `include` path, which allowed **3C** to rewrite the files with `itypes` based on its analysis of `icecast`'s use of them. The annotations to these headers are included in the Phase 1 **3C** line count (5529), which is why it is larger than the other two programs. Root cause analysis during Phase 1 also identified a generic AVL tree implementation as a pervasive source of *wild* pointers (e.g., more than 100 affected per root cause). We annotated the AVL tree's header `avl.h` with an `itype`-based interface matching the types we expected it should have, but deferred porting the implementation avl.c until Phase 2. We did something similar for icecast's use of a generic struct to contain audio format-specific data in a plugin-style infrastructure.[10]

During Phase 2 we further ported code to added missing bounds annotations, and to use generic features (e.g., on icecast's thread wrapper library). Sometimes a struct's generic-type use did not admit a Checked C type, so we chose to write wrapper functions to mediate access to it. We likewise needed to create wrappers or other exceptional cases for unsupported uses of generics on function pointers and in macros. All of this led to large, but idiomatic, changes throughout icecast. We are turning around our experience with icecast to the Checked C compiler team—better support for generics would have led to a lot less manual work.

*tiny-bignum.* tiny-bignum-c[11] is a small library for arbitrary-precision ("bignum"), unsigned integer arithmetic, where a bignum is implemented as a struct with a single field having a fixed-sized array of uints. The port to Checked C required only one run of **3C**: 99% of its pointers were converted and all of its required bounds were inferred. Just a few corrections were needed. The porting process uncovered two spatial safety bugs. One was due to an off-by-one error: the code for converting a bignum to a string failed to account for the space needed by the NULL terminator. This problem was made evident when **boun3c** identified the intended bound of the function, but then Checked C's compiler rejected the callsites, which were passing buffers short one byte. The other bug was due to an incorrect loop bound in this function, shown below, which was made manifest by a failed run-time check.

```
void bignum_to_string(_Ptr<struct bn> n, _Array_ptr<char> str : count(nbytes), int nbytes) {
  int i = 0;
  // This condition ensures that we have space for another 2 characters.
  while (... && (nbytes > (i + 1)))  {
   // Buffer overflow of str: SPRINTF_FORMAT_STR can write 8 bytes, 6 more than accounted for.
   🐞 sprintf(&str[i], SPRINTF_FORMAT_STR, ...);
    i += (2 * WORD_SIZE); ...
} }
```

## 7 RELATED WORK

Our work is related to past work aiming to automatically retrofit C code to be safe, and to work on automated type migration, in the gradual typing literature.

*Automated Analysis for Making C Safe.* Several prior works statically analyze C programs with the goal of ensuring memory safety. Many techniques use a simple compile-time analysis to instrument programs with run-time checks, but these can add significant overhead [Duck and Yap 2016; Kendall 1983; Nagarakatte et al. 2009; Serebryany et al. 2012; Steffen 1992]. CCured [Necula et al. 2005] aims to reduce overhead by adding bounds checks only where needed. Like **typ3c**, it employs a whole-program static analysis to identify *wild* and safe pointer uses, distinguishing pointers to arrays from pointers to single objects. But rather than output updated code, CCured compiles the analyzed C program directly. During compilation, CCured's *wild* pointers point to extra metadata; casts to/from wild pointers are therefore not allowed, resulting in an increase in wild pointers compared to **3C** (see Table 4). Array pointers in CCured are made "fat" with attached bounds information set at allocation time; as such, their static analysis need not infer bounds information at usage sites the way **boun3c** does. And since CCured is a compiler and not a source-to-source translator, it need not be concerned with the reusability (and re-analyzability) of its output. A follow-on to CCured called Deputy [Condit et al. 2007] includes Checked C-like bounds annotations on function and `struct` definitions, but these must be added by hand; a CCured-style analysis *within* a function (which introduces local fat pointer bounds) takes them into account [Zhou et al. 2006].

[10] Essentially, this is using void * to implement existential types.

[11] Available at https://github.com/kokke/tiny-bignum-c.

**typ3c** takes inspiration from algorithms for *type qualifier inference*, such as CQual [Foster et al. 2006]: the *chk* vs. *wild* distinction, and the *ptr*, *arr*, and *ntarr* checked types can be viewed as type qualifiers. **typ3c**'s *kind* constraints and solving algorithm are quite different from CQual's, due to wildness localization, and **typ3c** solves ptyp qualifier constraints using a novel algorithm that improves generality. CQual was also unconcerned with actually rewriting programs, which constrains some aspects of **3C**; e.g., **3C** cannot treat pointer types context sensitively because Checked C does not.

Cascade [Vakilian et al. 2015] uses a qualifier checker in a guess-and-check fashion to infer annotations, and speculates possible resolutions to "qualifier incompatibilities, which are caused by mismatches in the actual and expected qualifiers of expressions," from which the user can select one. **typ3c**'s notion of root cause is a specific kind of incompatibility, arising from a syntactic code pattern that generates a direct $wild \rightarrow q$ constraint (where $q$ is a pointer qualifier); such an edge arises due to an unsafe cast, invocation of a `void*` function, etc. Resolution of this problem will require changing the inducing code, directly, and doing so may have a positive downstream effect, which the root-cause analysis estimates. Cascade aims to handle qualifier incompatibilities from a generic lattice for which there are many possible annotation-based solutions, so they require a more involved (and more expensive) approach; its guess-and-check strategy is very unlikely to scale. We could imagine future work to **3C** that suggests changes and speculatively computes their impact, as Cascade does; we believe the main technical challenge will be making this speculation performant on large codebases.

Ruef et al. [2019] previously considered the problem of automatically converting a C program to Checked C. Their tool—call it CCC—is essentially a simpler version of **typ3c**; it lacks root-cause analysis and **boun3c**'s bounds inference. CCC could distinguish *ptr* types from array types, but did not distinguish *arr* from *ntarr*, and did not actually rewrite array types (just left a comment). CCC introduced the idea of using itypes or casts when callers/callees are not equally safe. **3C** also utilizes this idea, which we call *localizing wildness* (see Section 4.1). However, CCC 's unification-based inference algorithm, based loosely on CCured's algorithm, turns out to be unsound, for two reasons. First, it did not implement proper qualifier inference and as a result its implied notion of qualifier lattice—$q \in \{ptr, arr, wild\}$ with $ptr \sqsubset arr \sqsubset wild$—worked to localize wildness for function arguments but reversed the subtyping relationship that an *arr* can be used where a *ptr* is expected. As a result CCC could infer a result where a caller passed a *ptr* to a function expecting an *arr*. This problem is what led us to using paired qualifiers $(k, p)$ for **typ3c**, where $k \in \{chk, wild\}$ while $p \in \{ntarr, arr, ptr\}$ (Section 4.3). In addition, CCC 's localization of wildness only worked for function arguments/parameters, not returns, for which it was unsound. This is because it failed to appreciate that localizing wildness happens at the caller/callee boundary, regardless of the directionality of flow; instead, its constraints localized based on flow direction, so the handling of returns was backwards from what it should have been. **typ3c** solves this problem using reverse-flow edges for qualifiers $k$, and paired in/out nodes.

We are aware of no prior work that infers an array's bounds in terms of variables present in the program, as **boun3c** does. Numeric static analysis, e.g., as part of classical abstract interpretation [Cousot and Cousot 1977], is a longstanding area with many recent advances [Gange et al. 2015; Redini et al. 2019; Rodrigues et al. 2013]. We found these techniques not to scale that well (when used interprocedurally), and did not always infer bounds in terms of an expression over in-scope variables. That said, they can conceivably support more interesting bounds, such as `n+m+1` (not just `n` or `32`). It may be that **boun3c**'s approach could be enhanced with these techniques.

Rodrigues et al. [2019] address the related problem of generating in-bounds inputs for arguments to test functions using numeric static analysis techniques.

**boun3c**'s algorithm takes inspiration from *correlation analysis*. Locksmith [Pratikakis et al. 2011] introduced this kind of analysis to consistently correlate a pointer to potentially shared memory with the mutex that guards it. Correlations in Locksmith are inferred at dereference sites and propagated context sensitively. **boun3c**'s propagation phase is similar, but correlations are seeded at allocation sites or are guessed from context, and propagation must ensure bounds-flow respects variable scope, which Locksmith was not concerned with.

*Automated Type Migration.* A language with a *gradual type system* permits mixing statically and dynamically typed code—static type annotations are added incrementally to increase confidence in safety [Greenman and Felleisen 2018; Siek and Taha 2007; Tobin-Hochstadt et al. 2017]. Checked C is similar in spirit: the base language (C) is unsafe, and safety-enhancing annotations can be added incrementally. In this view, **3C** and our process for using it address the problem of *automated type migration* [Phipps-Costin et al. 2021], which is *how to automatically infer (or improve) absent static type annotations?* **typ3c**'s use of unification and subtyping constraints mirrors that of prior algorithms for automated type migration [Migeed and Palsberg 2020; Rastogi et al. 2012; Siek and Vachharajani 2008]. However, prior work mostly considers the effect of a single migration, and not the work that is left to complete a fuller port. We designed **3C** to be used within an iterative, human-in-the-loop process; e.g., *localized wildness* (Section 4.1) and *root cause analysis* (Section 4.2) specifically aim to reduce human effort, as does the use of heuristic bounds inference (Section 5.4). Future research may explore whether these ideas, and the two-phase process we designed to use them, can also work for other instances of automated type migration.

## 8 CONCLUSIONS AND FUTURE WORK

**3C** is a tool for providing automated assistance to a developer converting a C program to Checked C, a C-language extension that adds new *checked pointer* types whose use can ensure spatial memory safety. **3C**'s **typ3c** algorithm converts legacy pointers to checked ones using a variant of static type qualifier inference; **typ3c**'s novelty is in how constraints are generated and solved so to provide more general, localized results—including root causes of unsafety—to assist a developer using **3C** to interactively refactor a codebase. **3C**'s **boun3c** algorithm infers bounds annotations for checked array pointers, using a novel analysis to correlate pointers with potential in-scope bounds expressions. Experimental results on 11 programs totaling 319KLoC show **3C** to be effective at inferring checked pointer types; able to infer bounds annotations for roughly 3/4 checked array pointers; and supportive of an iterative workflow, able to complete much of the required annotation work of a full port to Checked C.

# A APPENDIX

## A.1 Analysis of *wild* pointers

Although **typ3c** was able to infer more checked pointers than other techniques, there are still a considerable number (32.1%) of pointers left as *wild*. Table 9 shows the detailed categorization of the *wild* pointers. First, these arise because of relatively fewer *root-cause pointers*—only 2,333 (8.8%) of all 8,489 *wild* are root causes. As mentioned in Section 4.2, the root-cause pointers are those pointers that are directly used in an unsafe manner and consequently have a direct edge from *wild*. It is interesting to see that many of the root causes arising for the same reasons; there were 91 unique reasons in total ($\mathbf{U}_r$). The top two reasons for *wild*ness for each program, and their total impact, are also given in the Table 9. The main reasons for **typ3c** marking pointers as *wild* are:

- **Invalid Cast**: Checked C does not allow casting checked pointers between incompatible pointer types, The most common was casts to/from `void*`; Checked C supports some forms of generic types, but **3C** does not attempt to infer them. For `vsftpd` and `lua`, there are assignments between pointer types whose compatibility cannot be verified statically; manual adjustments are required. In `lua`, these are casts from an internal string struct type to `char*` while in `vsftpd` the casts are from structs defined in system headers to private structs defined by `vsftpd`.

Table 9. Categorization of the *wild* pointers (**typ3c**$_w$) in each benchmark programs along with the underlying root-cause pointers (**Total**$_d$) along with associated with unique reasons (**U**$_r$)—the top two reasons with their impact on *wild* pointers are given.

| Program | Total Pointers (*TP*) | Wild Pointers (*wild*) (% of *TP*) | | | |
|---|---|---|---|---|---|
| | | typ3c$_w$ | Root Cause | | |
| | | | Total$_d$ | U$_r$ | Top two Reasons (% of typ3c$_w$) |
| vsftpd | 1,765 | 429 (24.3%) | 218 (12.4%) | 9 (0.5%) | Invalid Cast (64.37%)<br>Default void* type (16.9%) |
| icecast | 2,682 | 887 (33.1%) | 337 (12.6%) | 10 (0.4%) | Source code in non-writable file. (28.97%)<br>Default void* type (24.2%) |
| lua | 4,176 | 1,395 (33.4%) | 308 (7.4%) | 8 (0.2%) | Union field encountered (76.1%)<br>Invalid Cast (10.83%) |
| olden | 832 | 111 (13.3%) | 13 (1.6%) | 6 (0.7%) | Default void* type (56.8%)<br>Assigning from 0 depth pointer to 1 depth pointer. (36.4%) |
| parson | 686 | 179 (26.1%) | 99 (14.4%) | 8 (1.2%) | Inferred conflicting types (62.27%)<br>Invalid Cast (12.85%) |
| ptrdist | 920 | 236 (25.7%) | 23 (2.5%) | 15 (1.6%) | Unsafe call to allocator function. (45.63%)<br>Unchecked pointer in parameter (35.58%) |
| zlib | 647 | 262 (40.5%) | 48 (7.4%) | 6 (0.9%) | Default void* type (51.06%)<br>Invalid Cast (32.22%) |
| libtiff | 3,478 | 1,367 (39.3%) | 337 (9.7%) | 11 (0.3%) | Invalid Cast (52.02%)<br>Union field encountered (13.84%) |
| libarchive | 10,269 | 3,427 (33.4%) | 897 (8.7%) | 10 (0.1%) | Invalid Cast (58.89%)<br>Default void* type (18.75%) |
| thttpd | 829 | 195 (23.5%) | 53 (6.4%) | 7 (0.8%) | Default void* type (51.13%)<br>Source code in non-writable file. (27.57%) |
| tinybignum | 129 | 1 (0.8%) | 0 (0.0%) | 1 (0.8%) | Source code in non-writable file. (100.0%) |
| **Total** | 26,413 | 8,489 (32.1%) | 2,333 (8.8%) | 91 (0.3%) | |

- **Default** `void*` **type**: By default `void` pointers are not converted to checked pointers because the kind of conversion is unclear—it could be a `ptr<void>`, but that is less preferred than a generic type, when possible (Section 6.6).
- **External global variable**: Some C standard headers declare global pointer type variables. Checked C supports itypes on global variables to allow these pointers to be used in checked code, but itypes are not declared for all such pointers. There are also external functions which are *defined* in header files, e.g., `realpath` (in `stdlib.h`). The return values and parameters of these functions will be considered *wild* since the file is unwritable. Hence any pointer passed as an argument or assigned from these functions will be considered *wild*.
- **Unable to rewrite**: There are various reasons that **3C** might be unable to rewrite a pointer to a checked type. Most often, this happens when the pointer is declared inside macros or anonymous structs. To generate code that type checks, these pointers must be treated as *wild*.
- **Bad pointer type solution**: It is possible for contradictions to arise in the pointer type constraints generated with **typ3c**, most often due to unsafe uses of itype-annotated library functions. When this happens, the pointer cannot be meaningfully given a checked pointer type, so it remains unchecked.
- **Union or external struct**: Similarly to global variables, record types are often declared in header files. If the pointer field in the record has an unchecked type (and no itype), it cannot be changed, so it remains unchecked.

We found that these root-cause reasons provide meaningful insights into the failure modes of **typ3c**, and were helpful for driving the refactoring of the original code (Section 3).

Table 10. Pointers in each benchmark program (*not macro-expanded*) inferred by **typ3c** to be *chk* vs. previous work (CCured), and *wild* (**typ3c**$_w$); for the latter, the underlying root-cause pointers (**Total**$_d$) are associated with unique reasons (**U**$_r$)—the top two reasons with their impact on *wild* pointers are given, and column **Re**$_I$ is checked if the rewriter infrastructure is to blame.

| Program | Total Pointers (*TP*) | Checked pointers (*chk*) (% of *TP*) | | | Wild Pointers (*wild*) (% of *TP*) | | | | |
|---|---|---|---|---|---|---|---|---|---|
| | | | | | | Root Cause | | | |
| | | typ3c | typ3c$^f$ | CCured | typ3c$_w$ | Total$_d$ | U$_r$ | Top two Reasons (% of typ3c$_w$) | Re$_I$ |
| vsftpd$_m$ | 1,765 | **1,335 (75.6%)** | 1,226 (69.5%) | 999 (56.6%) | 430 (24.4%) | 218 (12.4%) | 9 (0.5%) | Invalid Cast (66.55%)<br>Default void* type (14.98%) | |
| icecast$_m$ | 2,682 | **1,794 (66.9%)** | 1,669 (62.2%) | 1,377 (51.3%) | 888 (33.1%) | 337 (12.6%) | 10 (0.4%) | Source code in non-writable file. (27.4%)<br>Default void* type (26.55%) | |
| lua$_m$ | 3,999 | **2,040 (51.0%)** | 1,713 (42.8%) | 1,484 (37.1%) | 1,959 (49.0%) | 503 (12.6%) | 9 (0.2%) | Pointer in Macro (28.02%)<br>Invalid Cast (25.91%) | |
| olden$_m$ | 832 | **634 (76.2%)** | 621 (74.6%) | 614 (73.8%) | 198 (23.8%) | 13 (1.6%) | 10 (1.2%) | Invalid Cast (66.67%)<br>Unsafe call to allocator function. (25.0%) | |
| parson$_m$ | 686 | **484 (70.6%)** | 351 (51.2%) | 291 (42.4%) | 202 (29.4%) | 99 (14.4%) | 8 (1.2%) | Inferred conflicting types (60.67%)<br>Source code in non-writable file. (16.78%) | |
| ptrdist$_m$ | 920 | **560 (60.9%)** | 537 (58.4%) | 508 (55.2%) | 360 (39.1%) | 23 (2.5%) | 18 (2.0%) | Invalid Cast (50.0%)<br>Default void* type (50.0%) | |
| zlib$_m$ | 647 | **146 (22.6%)** | 141 (21.8%) | 106 (16.4%) | 501 (77.4%) | 57 (8.8%) | 7 (1.1%) | Pointer in Macro (68.19%)<br>Invalid Cast (18.85%) | |
| libtiff$_m$ | 3,448 | **1,935 (56.1%)** | 1,850 (53.7%) | 1,176 (34.1%) | 1,513 (43.9%) | 361 (10.5%) | 12 (0.3%) | Invalid Cast (78.34%)<br>Default void* type (7.23%) | |
| libarchive$_m$ | 10,261 | **6,785 (66.1%)** | 6,158 (60.0%) | 4,918 (47.9%) | 3,476 (33.9%) | 937 (9.1%) | 11 (0.1%) | Invalid Cast (58.19%)<br>Default void* type (20.01%) | |
| thttpd$_m$ | 829 | **631 (76.1%)** | 615 (74.2%) | 448 (54.0%) | 198 (23.9%) | 54 (6.5%) | 8 (1.0%) | Default void* type (46.97%)<br>Source code in non-writable file. (26.91%) | |
| tinybignum$_m$ | 129 | **128 (99.2%)** | 117 (90.7%) | 117 (90.7%) | 1 (0.8%) | 0 (0.0%) | 1 (0.8%) | Source code in non-writable file. (100.0%) | |
| **Total** | 26,198 | **16,472 (62.9%)** | 14,998 (57.2%) | 12,038 (46.0%) | 9,726 (37.1%) | 2,602 (9.9%) | 103 (0.4%) | | |

## A.2 3C on non-macro expanded programs

In this section, we report the effectiveness of **typ3c** when run on the original, non macro-expanded benchmarks. Recall (Section 6.1) that we expanded macro uses (but retained all preprocessor directives and macro definitions) since they oftentimes prevented a safe source code rewriting. Table 10 shows the results of running **typ3c**, **typ3c**$^f$, and, **CCured** on non-preprocessed programs. The table also shows the analysis of *wild* pointers similar to Appendix A.1.

The detection rate of *chk* pointers only slightly decreased to 62.9% from 67.9%. The main reason for this is the limitation of `clang`tool-based rewriting library that **3C** uses to output the source code with pointer annotations. This library does not support rewriting statements inside preprocessor declarations (e.g., `#define`). Thus, **typ3c** is forced to consider all pointers within the corresponding statements to be *wild*, since they could not be rewritten. The affects the benchmarks which extensively make use of macros, as indicated by a check in column **Re**$_I$. In the case zlib$_m$, 68.19% of the *wild* pointers are because they depend on pointers declared in macros, which zlib$_m$ extensively uses to define functions. lua$_m$ is similarly affected by macro use.

## A.3 3C vs. least vs. greatest solving algorithm

To determine the type of checked pointers, **typ3c** uses a novel three-step solving algorithm for ptyp constraints.

*Bounded returns.* One detail left out of the algorithm description in Section 4.3 is that we use the least solution for a return variable $q_{ret}$ only if $q_{ret}$ is *bounded*, meaning that it is constrained, directly or transitively, by a literal (e.g., *arr* from array usage, or from interacting with an itype as is happening with `malloc`), or an already-solved parameter qualifier; otherwise we choose the greatest solution. This is because we find that a completely unconstrained return is typically a singleton pointer, and hence should solve to $ptr$. Consider this example:

Table 11. Breakdown of checked pointer types inferred by **typ3c**; its three-step solving approach (**typ3c**) is contrasted with *Greatest* and *Least* solutions, where differences between these and **typ3c** are highlighted in red.

| Program | Total | Checked Pointers (*chk*) | | | | | | | | |
|---|---|---|---|---|---|---|---|---|---|---|
| | | ptr (% of Total) | | | arr (% of Total) | | | ntarr (% of Total) | | |
| | | typ3c | Greatest | Least | typ3c | Greatest | Least | typ3c | Greatest | Least |
| vsftpd | 1,336 | 1,199 (89.7%) | 1,199 (89.7%) | 502 (37.6%) | 44 (3.3%) | 44 (3.3%) | 638 (47.8%) | 93 (7.0%) | 93 (7.0%) | 196 (14.7%) |
| icecast | 1,795 | 1,429 (79.6%) | 1,429 (79.6%) | 848 (47.2%) | 54 (3.0%) | 54 (3.0%) | 406 (22.6%) | 312 (17.4%) | 312 (17.4%) | 541 (30.1%) |
| lua | 2,781 | 2,273 (81.7%) | 2,273 (81.7%) | 1,572 (56.5%) | 254 (9.1%) | 254 (9.1%) | 622 (22.4%) | 254 (9.1%) | 254 (9.1%) | 587 (21.1%) |
| olden | 721 | 571 (79.2%) | 573 (79.5%) | 399 (55.3%) | 130 (18.0%) | 128 (17.8%) | 212 (29.4%) | 20 (2.8%) | 20 (2.8%) | 110 (15.3%) |
| parson | 507 | 405 (79.9%) | 405 (79.9%) | 116 (22.9%) | 9 (1.8%) | 9 (1.8%) | 277 (54.6%) | 93 (18.3%) | 93 (18.3%) | 114 (22.5%) |
| ptrdist | 684 | 465 (68.0%) | 466 (68.1%) | 246 (36.0%) | 181 (26.5%) | 180 (26.3%) | 258 (37.7%) | 38 (5.6%) | 38 (5.6%) | 180 (26.3%) |
| zlib | 385 | 293 (76.1%) | 293 (76.1%) | 51 (13.2%) | 86 (22.3%) | 86 (22.3%) | 198 (51.4%) | 6 (1.6%) | 6 (1.6%) | 136 (35.3%) |
| libtiff | 2,111 | 1,694 (80.2%) | 1,694 (80.2%) | 281 (13.3%) | 177 (8.4%) | 177 (8.4%) | 1,220 (57.8%) | 240 (11.4%) | 240 (11.4%) | 610 (28.9%) |
| libarchive | 6,842 | 5,532 (80.9%) | 5,532 (80.9%) | 2,534 (37.0%) | 896 (13.1%) | 896 (13.1%) | 2,571 (37.6%) | 414 (6.1%) | 414 (6.1%) | 1,737 (25.4%) |
| thttpd | 634 | 341 (53.8%) | 341 (53.8%) | 96 (15.1%) | 57 (9.0%) | 57 (9.0%) | 163 (25.7%) | 236 (37.2%) | 236 (37.2%) | 375 (59.1%) |
| tinybignum | 128 | 110 (85.9%) | 110 (85.9%) | 55 (43.0%) | 3 (2.3%) | 3 (2.3%) | 54 (42.2%) | 15 (11.7%) | 15 (11.7%) | 19 (14.8%) |
| **Total** | 17,924 | 14,312 (79.8%) | 14,315 (79.9%) | 6,700 (37.4%) | 1,891 (10.6%) | 1,888 (10.5%) | 6,619 (36.9%) | 1,721 (9.6%) | 1,721 (9.6%) | 4,605 (25.7%) |

```
int *foo(void) { return (int *)0; }
```

The returned value is unbounded, and if given a least solution would return *ntarr*, which is probably not what we wanted.

*Comparing solving algorithms.* We evaluated the benefit of this novel solving algorithm against reasonable alternatives. Table 11 shows its solution in comparison with the least, and greatest, solutions. The percentages in red identify the difference with **typ3c**'s solution.

For all the benchmarks, the least solution significantly diverges (by 42% or so) from the **typ3c** solution. This solution tends to produce many more *ntarr* pointers, which has the downsides discussed in Section 6.3.2.

We were surprised to see that the greatest solution is a very close match to **typ3c**'s results for the six benchmark programs—the differences are with `olden` and `ptrdist` by a few pointers. This makes sense on reflection. The **typ3c** solving algorithm's first step is to compute the greatest solution for parameters, so their solutions should match the overall-greatest solution. For locals, and when returns have no bound, **typ3c** also uses the greatest solution. When returns are bounded, **typ3c** uses the least solution, which has a downstream effect on some locals; these are the only opportunities for a difference. But the difference will only manifest in cases that a returned pointer is never used at its most specific (i.e., least) type. However, this situation does not arise much in our benchmarks. Moreover, it will only come up for pointer-typed returns, of which there are fewer than pointer-typed parameters. Nonetheless, libraries that return array pointers should benefit, as would an incremental style of development that runs **3C** on a single file without considering its callers at the same time.

## A.4 boun3c auxiliary functions

Algorithm 3 shows the pseudocode of the auxiliary function used in Algorithm 3 (Section 5).

**Algorithm 3: boun3c** bounds inference - Auxiliary Functions.

**Function** `GetBoundsFlow`($c, SB_n$):
- $SB \leftarrow \emptyset$
- **forall** $(\vartheta_i, s_i) \in SB_n$ **do**
  - $F_i \leftarrow sfg$.`Reachable`(vis($ns(c)$), $s_i$)
  - $SB \leftarrow SB \cup \{(\vartheta_i, x) \mid \forall x \in F_i\}$
- **return** $SB$

**Function** `GetCommonBoundsSet`($SSB_c$):
- $\{SB_1, ..., SB_n\} \leftarrow SSB_c$
- **return** $\{b \mid (b \in SB_1) \land ... \land (b \in SB_n)\}$

**Function** `FindBounds`($c, \beta, CB$):
- $SB_c \leftarrow \emptyset$
- **if** $c \in \beta$ **then**
  - $SB_c \leftarrow \{\beta(c)\}$
- **else if** $c \in CB$ **then**
  - $SB_c \leftarrow CB(c)$
- **return** $SB_c$